# Visible-to-ultraviolet (<340 nm) photon upconversion by triplet–triplet annihilation in solvents[†]

Yoichi Murakami,[1,*] Ayumu Motooka,[1‡] Riku Enomoto,[1] Kazuki Niimi,[2] Atsushi Kaiho[2] and Noriko Kiyoyanagi[2]

[1] School of Engineering, Tokyo Institute of Technology, 2-12-1 Ookayama, Meguro-ku, Tokyo 152-8552, Japan.

[2] Nippon Kayaku Co., Ltd., 3-31-12 Shimo, Kita-ku, Tokyo 115-8588, Japan.

*E-mail: murakami.y.af@m.titech.ac.jp

[†] Electronic Supplementary Information (ESI) available.

[‡] Present address: FANUC Corp., Oshino-mura, Yamanashi 401-0597, Japan.

# Abstract

In this article, visible-to-ultraviolet photon upconversion (UV-UC) by triplet–triplet annihilation in the emission range shorter than 340 nm, which is previously unexplored, is presented and the relevant physicochemical characteristics are elucidated. Investigations were carried out in several deaerated solvents using acridone and naphthalene derivatives as a sensitizer and emitter, respectively. Both upconversion quantum efficiency and sample photostability under continuous photoirradiation strongly depended on the solvent. The former dependence is governed by the solvent polarity, which affects the triplet energy level matching between the sensitizer and emitter because of the solvatochromism of the sensitizer. To elucidate the latter, first we investigated the photodegradation of samples without the emitter, which revealed that the sensitizer degradation rate is correlated with the difference between the frontier orbital energy levels of the sensitizer and solvent. Inclusion of the emitter effectively suppressed the degradation of the sensitizer, which is ascribed to fast quenching of the triplet sensitizer by the emitter and justifies the use of ketonic sensitizers for UV-UC in solvents. A theoretical model was developed to acquire insight into the observed temporal decays of the upconverted emission intensity under continuous photoirradiation. The theoretical curves generated by this model fitted the experimental decay curves well, which allowed the reaction rate between the emitter and solvent to be obtained. This rate was also correlated with difference between the frontier orbital energy levels of the emitter and solvent. Finally, based on the acquired findings, general design guidelines for developing UV-UC samples were proposed.

# 1. Introduction

Photon upconversion (UC) is a technology to convert presently wasted sub-bandgap photons into those with higher energies (i.e., light of shorter wavelength), which are useful in many fields including photovoltaics and photocatalysis. To date, UC using triplet–triplet annihilation (TTA) between organic molecules has been widely explored because of its applicability to low-intensity and non-coherent light.[1–5] Most of the previous studies focused on visible-to-visible UC.[1–40] If TTA-UC technology can be reliably extended to the ultraviolet (UV) region (<400 nm), it will become suitable for a broader range of applications, such as for hydrogen generation by water splitting using anatase titanium dioxide ($a$-TiO$_2$), which has a band gap of 3.2 eV ($\lambda_{\text{gap}}$ ~385 nm).[41]

Since the pioneering studies by Castellano and co-workers[42,43] and Merkel and Dinnocenzo,[44] there have been multiple reports[45–52] exploring UC of visible light to UV light (UV-UC). Here, the principle of TTA-UC is briefly described (Fig. 1a). First, a sensitizer molecule absorbs a low-energy photon (visible photon in this context) and transforms to the excited singlet (S$_1$) state, which immediately converts to the triplet (T$_1$) state with a certain quantum yield through intersystem crossing. If the energy of the T$_1$ state of the emitter is similar to or lower than that of the sensitizer, the T$_1$ energy of the sensitizer can be transferred to the emitter (triplet energy transfer; TET), creating a T$_1$ emitter (Fig. 1b). When two T$_1$ emitters interact and undergo TTA, an S$_1$ emitter can be generated from which an upconverted photon (UV photon in this context) is emitted as delayed fluorescence.

Most previous UV-UC studies were carried out using pyrene or a derivative, whose UC emission maxima range between ca. 375 and 425 nm,[42,45,50] or 2,5-diphenyloxazole (PPO), whose UC emission maxima range between 350 and 400 nm,[43,44,46,48,49,51] as the emitter. For PPO, 2,3-butanedione (biacetyl) has often been used as the sensitizer.[43,46,49] As far as we surveyed, except for our previous technical documents[53] on which this study is based, the shortest emission peak wavelength reported for UV-UC by TTA is 343–344 nm using terphenyl as the emitter.[47,50] Therefore, UV-UC with emission maxima shorter than 340 nm has not been well explored thus far.

Shortening emission wavelengths further is meaningful for the following reasons. First, although $\lambda_{gap}$ of $a$-TiO$_2$ is ca. 385 nm, which was determined by tangentially extrapolating its

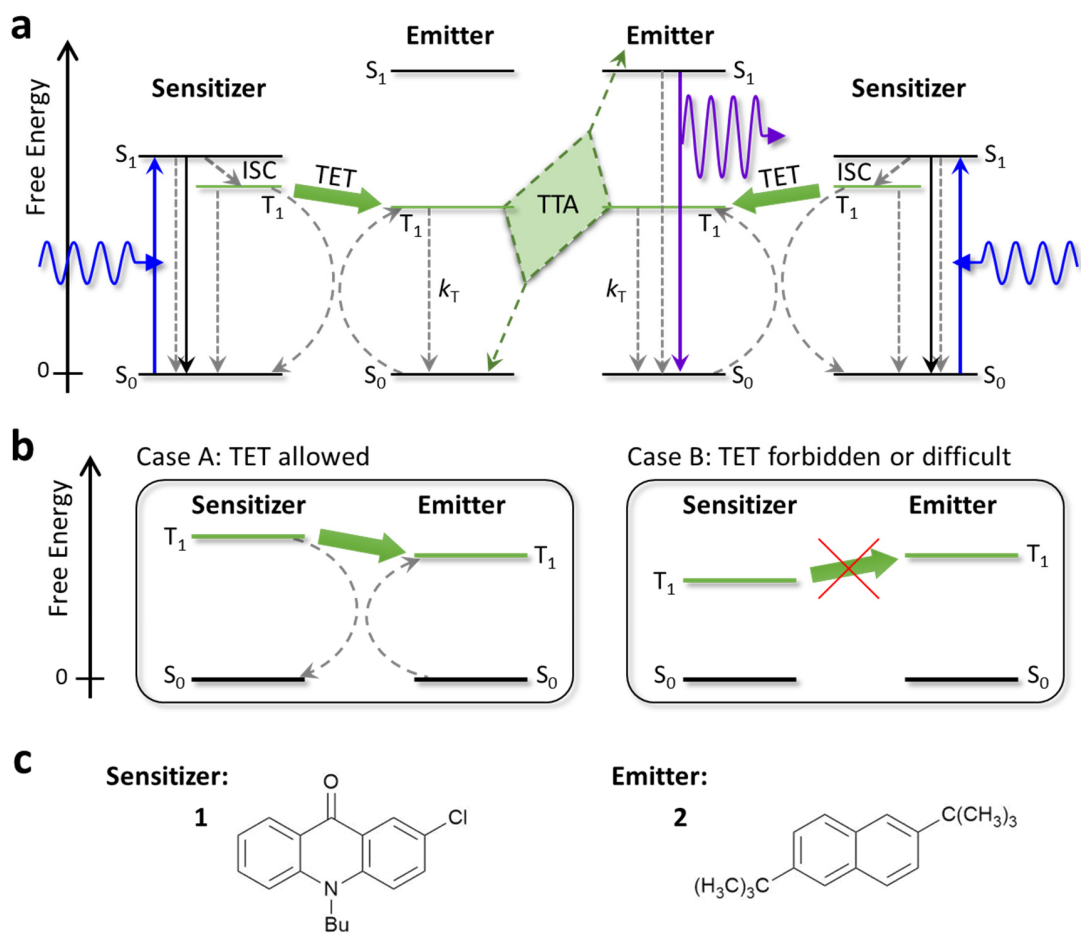

**Fig. 1** (a) Schematic diagram of the process of TTA-UC. ISC, TET, and TTA mean intersystem crossing, triplet energy transfer, and triplet–triplet annihilation, respectively. Solid and dashed arrows represent radiative and non-radiative processes, respectively. (b) Schematic depictions of two cases where TET is allowed (Case A) and forbidden or difficult (Case B). (c) Molecular structures of the sensitizer **1** and emitter **2** used in this study.

absorbance or reflectance spectrum to the horizontal axis,[54] a general characteristic of semiconductors is that their absorption coefficient is low near $\lambda_{gap}$.[55] For example, sufficient absorption is attained only below ca. 350 nm in the case of *a*-TiO$_2$ nanoparticles.[54,56] Second, the quantum efficiency of water-splitting photocatalysts increases with the energy of incident photons.[57] This present article investigates UV-UC with emission maxima shorter than 340 nm and elucidates the relevant physicochemical characteristics.

However, we have noticed that such UV-UC, whether the samples used in this article or other samples such as those made using biacetyl and/or PPO, is accompanied by non-trivial or sometimes remarkable photodegradation, although such characteristics were not explicitly presented and discussed previously. Only recently, Lee et al.[50] showed fast photodegradation caused by continuous photoirradiation at 455 nm in deaerated tetrahydrofuran (THF) when PPO and terphenyl were used as emitters. They showed that, among the emitters tested, only pyrene exhibited satisfactory photostability in deaerated THF.[50]

Previously, we reported visible-to-visible UC in systems using an ionic liquid as the solvent.[16,21–23,28] These samples, when properly sealed, exhibited excellent photostability and their lifetime exceeded several years (Fig. S1, ESI†). However, when the same ionic liquid was combined with the sensitizer and emitter used in the present study for UV-UC (Fig. 1c), such photostability was not observed (Fig. S1, ESI† and also below). We also found that the combination of biacetyl and PPO in deaerated dimethylformamide (DMF), which were used previously,[46,49] showed poor stability under continuous photoirradiation (Fig. S2, ESI†).

Based on these observations, we consider that UV-UC at wavelengths shorter than ca. 370 nm tends to suffer from low photostability, presumably because the use of high-energy triplet states may induce photochemical reactions, such as hydrogen abstraction from the solvent. This is an unaddressed issue that should be investigated before UV-UC technology is used in applications. Therefore, it is important to obtain understanding of the governing factors and/or mechanism of such photodegradation in UV-UC.

In this study, based on our previous technological findings regarding UV-UC,[53] we develop UV-UC samples that exhibit photoemission peaks in the 320–340 nm range. We find that both the UC quantum efficiency ($\Phi_{UC}$) and photostability of these samples depend on the solvent. To understand this phenomenon, we conduct a systematic investigation by performing both experiments and theoretical analysis. The aim of this article is to elucidate the factors governing such solvent dependence and obtain general guidelines for designing UV-UC systems with high UC efficiency and photostability.

## 2. Experimental

We used 10-butyl-2-chloro-9(10H)-acridinone (**1**) and 2,6-di-*tert*-butylnaphthalene (**2**) as the sensitizer and emitter, respectively (Fig. 1c). Both **1** and **2** (purity: >98%) were purchased from TCI; **1** was recrystallized before use and **2** was used as received. We chose **1**, in which the photoexcitation is the n→π* transition, because the small overlap between the n and π* orbitals around its carbonyl group leads to a small $S_1$–$T_1$ energy gap and the n,π* state has a high quantum yield of $S_1$-to-$T_1$ intersystem crossing ($\Phi_{T,sen}$),[58] both of which are desirable for sensitizers for TTA-UC. After testing several acridones, we found that **1** was preferable over the other candidates because of its visible absorption in the 400–425 nm range (Fig. S3, ESI†) and ability to undergo

TET with naphthalenes. We chose **2** because of its relatively high fluorescence quantum yield and suitable fluorescence spectrum for the purpose of this study.

Samples were prepared using the solvents listed in Table 1. Details of the solvents are given in Table S1 in the ESI†. We included D-limonene because it has been reported to prevent degradation of solutes in visible-to-visible UC by functioning as a strong antioxidant that quickly scavenges residual oxygen.[59] Additionally, in the former half of this study, we included the ionic liquid [C$_4$dmim][NTf$_2$] as a reference solvent because it enables highly stable red-to-blue UC[16,21–23] (Fig. S1, ESI†). Throughout this report, the concentrations of **1** and **2** were $2\times10^{-4}$ and $2\times10^{-3}$ M, respectively. A continuous-wave laser with an emission wavelength of 405 nm and spot diameter of 0.8 mm was used as the excitation source unless otherwise specified. The absorption spectra of **1** and **2** are depicted in Fig. S3 in the ESI† and their fluorescence spectra are shown in Fig. 2a.

**Table 1** List of samples and selected results

| Solvent $\langle E_T(30) \rangle^a$ $(P')^b$ | $A_{405nm}{}^f$ | $\Phi_{F,sen}$ | $\Phi_{T,sen}{}^g$ | $\Phi_{F,emi}$ | $\Phi_{UC}$ / % (Exct. Intensity / W cm$^{-2}$) | $k_{sen,degr}{}^j$ / M s$^{-1}$ | $k_{emi,rxn}{}^k$ / s$^{-1}$ |
|---|---|---|---|---|---|---|---|
| Hexane ⟨30.9⟩ (0.1) | 0.067 | 0.006 | 0.994 | 0.33 | 4.5 (0.40)$^h$<br>8.2 (1.75) | $7.35 \times 10^{-7}$<br>— | $5.73 \times 10^{-3}$ |
| Ethyl Acetate ⟨38.0⟩ (4.4) | 0.20 | 0.274 | 0.726 | 0.39 | 1.9 (0.20)$^h$<br>4.9 (1.74) | $2.35 \times 10^{-6}$<br>— | $1.92 \times 10^{4}$ |
| Toluene ⟨33.9⟩ (2.4) | 0.23 | 0.191 | 0.809 | 0.57 | 2.2 (0.17)$^h$ | $1.56 \times 10^{-5}$ | $4.47 \times 10^{-1}$ |
| Acetonitrile ⟨45.6⟩ (5.8) | 0.15 | 0.598 | 0.402 | 0.43 | 0.38 (0.50)$^h$ | $4.09 \times 10^{-7}$ | $3.29 \times 10^{4}$ |
| DMF[c] <43.2> (6.4) | 0.17 | 0.593 | 0.407 | 0.49 | 0.015 (0.44)[h] | $1.48 \times 10^{-5}$ | — |
| D-Limonene <N/A>[d] (N/A) | 0.20 | 0.030 | 0.970 | 0.16 | $\cong$ 0 (0.15)[h] | $1.95 \times 10^{-5}$ | — |
| [C$_4$dmim][NTf$_2$] <40.9>[e] (N/A) | 0.15 | 0.575 | 0.425 | 0.58 | 0.25 (0.44)[i] | $1.15 \times 10^{-6}$ | — |
| Methanol <55.4> (5.1) | 0.16 | 0.657 | 0.343 | 0.37 | $\cong$ 0 (0.52)[h] | — | — |

[a] From ref. 60 unless otherwise specified. [b] From ref. 61. [c] N,N-dimethylformamide. [d] Measured using Reichardt's dye but no absorption peak was found in the visible to near-infrared region. [e] Measured using Reichardt's dye. [f] Absorbance of **1** at 405 nm with an optical path length of 1 mm. [g] Determined by assuming the Ermolev's rule $\Phi_{T,sen} = 1 - \Phi_{F,sen}$. [h] Excitation intensity where the T$_1$ state of **1** is generated at $1.9 \times 10^{-3}$ M/s. [i] Excitation intensity where the T$_1$ state of **1** is formed at $1.77 \times 10^{-3}$ M/s. [j] Obtained by the procedure described in Section 11 of the ESI†. [k] Obtained from the fit to the experimental decay curves of the UC emission intensity using our model; see Fig. 4c.

Except for the sample with [C$_4$dmim][NTf$_2$], all samples, which contained both **1** and **2** or only **1**, underwent at least seven (typically eight or nine) freeze–pump–thaw (FPT) cycles using our FPT system to carefully remove dissolved air. Our FPT system consisted of a small glass jar (#33.010007.11A.710, EVAC) with a flange, an O-ring-sealed stainless-steel (SUS) flange coupling to it, and all-SUS vacuum line consisting of a flexible metal hose and Swagelok valves and tube fittings. The vacuum line was connected to an oil-free dry scroll pump (nXDS15i, Edwards) able to attain a vacuum of ca. 1 Pa. To efficiently remove dissolved gas, the volume of liquid in the jar was small (ca. 2 mL). The increase of solute concentration induced by the FPT cycles was negligible for all samples, as confirmed by their unchanged absorbance in UV-vis measurements. Typically, the emergence of bubbles in the liquid ended within three or four FPT cycles, and thus the aforementioned number of FPT cycles was believed to be sufficient. After the FPT cycles, the glass jar coupled with a closed SUS valve was detached from the vacuum line and transferred into a vacuum-type SUS glovebox containing freshly replaced nitrogen gas (purity: >99.998%). In the glovebox, the liquid sample was injected into a square glass capillary (inner dimensions: 1×1 mm, outer dimensions: 2×2 mm, length: 27 mm) with a closed end. The open top of the capillary was immediately closed with a low-melting-point solder as previously described.[16,21–23] The seals were checked by placing the capillary under vacuum for a long period (hours or days); an effective seal was confirmed by the sample volume remaining constant. This sealing method works for at least several years (e.g., the sample in Fig. S1, ESI†). For the sample with [C$_4$dmim][NTf$_2$], oxygen and moisture were removed by stirring the sample, which had a small volume (<600 μL), in an open vial at 60 °C and 200 rpm for 2 h inside a nitrogen-filled glovebox equipped with a gas-purification system (OMNI-LAB, VAC; oxygen and moisture: <1 ppm) before it was sealed in a glass capillary inside the glovebox.

Time-resolved measurements of the UC emission intensity were carried out using nanosecond light pulses generated from an optical parametric oscillator (OPO; NT-242, Ekspla) at 410 nm and 20 Hz. The fluorescence quantum yields of **1** and **2** ($\Phi_{F,sen}$ and $\Phi_{F,emi}$, respectively) and their fluorescence spectra were acquired by an absolute quantum yield spectrometer (Quantaurus-QY, Hamamatsu) using a quartz cell (1×1 cm); the solute concentration for these measurements (of the order of $10^{-5}$ to $10^{-4}$ M) was chosen so that the absorptance of each sample in the integrating sphere was between 0.35 and 0.55 at the excitation wavelength.

In reference experiments, photodegradation was controllably induced in a sample using a setup where the excitation laser beam was expanded to irradiate almost the entire volume of a sample liquid (ca. 2 mL) in a hermetically sealed glass vial from below (see Fig. S4 in the ESI† for details). In these experiments, the photoirradiation was continued until each molecule of **1** in the sample turned to the T$_1$ state 85 times on average. The duration of photoirradiation was set by assuming that the initial absorbance of **1** at 405 nm did not change during the course of the irradiation. All photoemission spectra in this report were corrected by the wavelength-dependent sensitivities of the grating in a monochromator and CCD array detector as described in our previous reports.[16,21–23]

All quantum-chemical simulations were carried out using Gaussian 16® at the B3LYP/6-31G++(d,p) level.

## 3. Results and discussion

The fluorescence and absorption spectra of sensitizer **1** exhibited large solvatochromic shifts whereas those of emitter **2** did not (see Fig. 2a for the fluorescence spectra and Fig. S3 in the ESI† for the optical absorption spectra). This behavior is ascribed to the large (negligible) permanent dipole moment of **1** (**2**) (Fig. S5, ESI†). Figure 2b shows photoemission spectra of samples

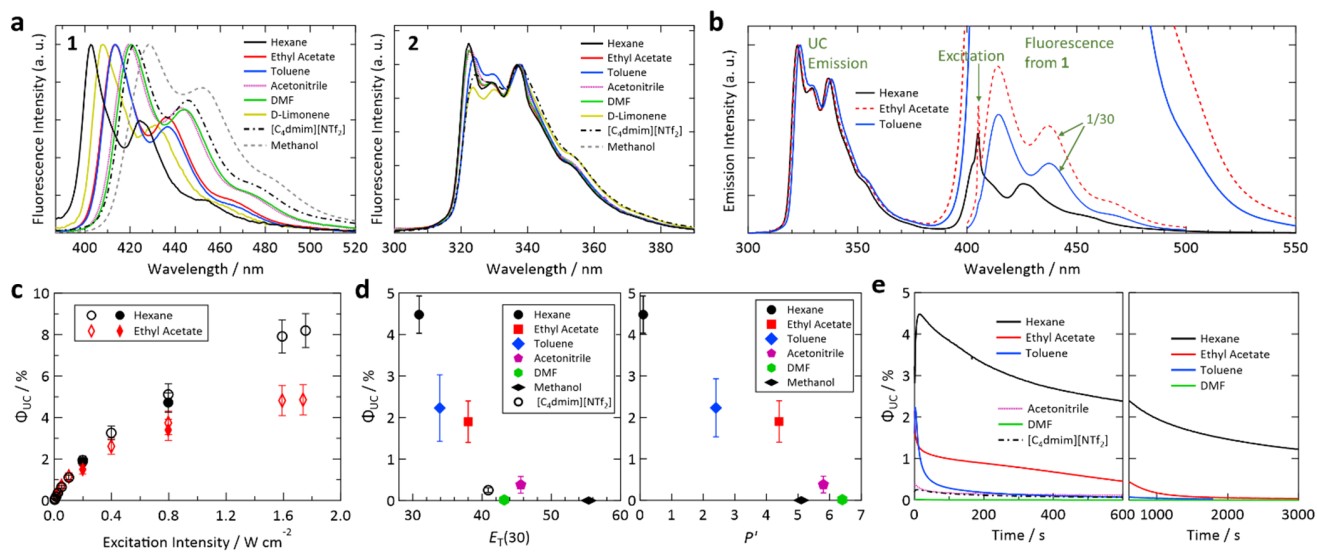

**Fig. 2** (a) Fluorescence spectra of **1** and **2** in the solvents used in this study. (2) Photoemission spectra of selected samples excited at 405 nm. (c) Dependence of the upconversion (UC) quantum efficiency of the samples prepared using hexane and ethyl acetate on excitation intensity. Open symbols denote data acquired while increasing excitation intensity and filled symbols represent data acquired with decreasing excitation intensity. (d) UC quantum efficiencies plotted against $E_T(30)$ (left) and solvent polarity scale $P'$ (right). (e) Temporal profiles of UC quantum efficiency, which is proportional to UC emission intensity, measured for samples prepared using different solvents under continuous photoirradiation at 405 nm. In (b), (d), and (e), the excitation light intensity was chosen such that the irradiation generated the $T_1$ state of **1** at a rate of $1.9\times10^{-3}$ M/s ($1.77\times10^{-3}$ M/s for [C$_4$dmim][NTf$_2$]). See the main text for the details.

prepared using hexane, ethyl acetate, and toluene upon excitation at 405 nm. The UC emission spectra were structured with the emission maximum at 322 nm and other peaks in the range of 320–340 nm, which are at shorter wavelengths than the spectra of previous UV-UC systems.[42–52] The photoemission spectra also contained peaks originating from fluorescence from the $S_1$ state of **1** in the 400–500 nm range. The intensity of this fluorescence relative to that of the UC emission varied considerably between samples, which is partially attributed to the difference of $\Phi_{F,sen}$ in these solvents ($\Phi_{F,sen}$ = 0.006, 0.274, and 0.191 in hexane, ethyl acetate, and toluene, respectively; cf. Table 1).

The dependence of $\Phi_{UC}$ of the samples with hexane and ethyl acetate on excitation intensity was determined (Fig. 2c). For $\Phi_{UC}$ in this article, we customarily describe efficiency in percent and thus the maximum is 100%, which is twice the maximum UC quantum yield of 0.5. The emission intensity between 310 and 380 nm was used to calculate $\Phi_{UC}$; i.e., the emission between 380 and 405 nm was not used to exclude the tail of the fluorescence and thermally induced UC emission. The procedure used to determine $\Phi_{UC}$ is described in Section 7 of the ESI†. As shown in Fig. 2c, the samples with hexane and ethyl acetate attained high $\Phi_{UC}$ of 8.2% and 4.9%, respectively, at an excitation intensity of ca. 1.75 W/cm². The data points in Fig. 2c were acquired while first increasing the excitation intensity and then while decreasing the excitation intensity to confirm the reproducibility of the $\Phi_{UC}$ values. Although $\Phi_{UC}$ measured while decreasing the excitation intensity were slightly lower than those obtained with increasing excitation intensity for both samples, the differences were smaller than the related error bars and thus $\Phi_{UC}$ values were considered reproducible.

We found that $\Phi_{UC}$ and photostability strongly depended on the solvent. To systematically compare $\Phi_{UC}$ and the rates of photoinduced changes of the samples prepared using different solvents, in the following experiments we set the laser power irradiated onto the sample sealed in a glass capillary (see the Experimental section for details) such that the irradiation generated the T$_1$ state of **1** at a rate of 1.9×10$^{-3}$ M/s in the photoirradiation volume (a cylinder with a diameter of 0.8 mm and length of 1 mm). At this rate, each **1** molecule in the volume transitions to the T$_1$ state 9.5 times per second. Note that the rate was 1.77×10$^{-3}$ M/s in the sample with [C$_4$dmim][NTf$_2$]. The actual laser power irradiated onto each sample, which was in the range of 0.73–2.6 mW or 0.15–0.52 W/cm$^2$, was determined using the absorbance at 405 nm ($A_{405nm}$) and $\Phi_{T,sen}$ listed in Table 1. The $\Phi_{T,sen}$ values were estimated assuming Ermolev's rule of $\Phi_T \approx 1 - \Phi_F$.[58]

The determined $\Phi_{UC}$ values are plotted against the polarity scales $E_T(30)$[60] and $P'$[61] in Fig. 2d. We were unable to determine the $E_T(30)$ value for D-limonene because it did not exhibit an absorption peak in the visible to near-infrared range in a solution of Reichardt's dye. No $P'$ values for D-limonene or [C$_4$dmim][NTf$_2$] were found in the literature. From these plots, we found that $\Phi_{UC}$ is correlated with the solvent polarity and decreases as the polarity increases.

As mentioned above, **1** has a large dipole moment (Fig. S5, ESI†) and thus exhibits a large bathochromic shift as the solvent polarity increases, whereas **2** does not (Fig. 2a). Therefore, as the solvent polarity increases, the T$_1$ level of **1** is considered to be lowered relative to that of **2**, making TET thermodynamically unfavorable (i.e., Case B in Fig. 1b). The solvent dependence of $\Phi_{UC}$ of our samples is mainly attributed to this mechanism. In addition, the solvent dependences of $\Phi_{T,sen}$ and $\Phi_{F,emi}$ (Table 1) should also affect $\Phi_{UC}$.

The stability of the samples under continuous photoirradiation strongly depended on the solvent (Fig. 2e). For example, UC emission in toluene decayed rapidly whereas that in hexane lasted much longer; the reason for this behavior is investigated below. It is noted that no UC emission was observed when D-limonene and methanol were used (Table 1). While the lack of UC emission in methanol can be explained by the above discussion regarding Fig. 2d, the reason for the absence of UC emission in D-limonene is unclear. It may be caused by the high reactivity of D-limonene, which has a reactive unsaturated C=C bond, with high-energy triplet states of **1** and **2**.

Here, we note the following three points. First, although the use of solvents with different certified purities resulted in a minor but recognizable effect on the intensity of UC emission, this difference did not alter the qualitative profile of the temporal UC emission intensity change (Fig. S6, ESI†). Second, the temporal decays of the UC emission intensity observed in Fig. 2e were not considered to be governed by residual oxygen in the solvents, which was the case in previous visible-to-visible UC studies.[62–66] This is partly because the use of D-limonene, which scavenged residual oxygen efficiently and helped to attain stable visible-to-visible UC,[59] completely suppressed the UC emission in the present study. That the UC emission decays observed in Fig. 2e were not caused by residual oxygen was also supported by the thorough FPT treatment and tightly sealed samples used here. Third, the decay rate of the UC emission in hexane in the present study is much slower than that of a previously reported biacetyl/PPO/DMF system[46,49] when compared using the similar triplet generation rate on **1** (Fig. S2, ESI†).

In the following investigations, we excluded the sample with methanol because it did not realize UC and the sample with [C$_4$dmim][NT$_2$] because its UC efficiency was low and the photochemical reaction with a molten salt is complex.

To understand the solvent-dependent photostability of our samples, first, we investigated samples containing only **1**. When each sample in a glass capillary was excited at a triplet generation rate of $1.9 \times 10^{-3}$ M/s, the decay rate of the fluorescence intensity of **1** depended on the type of solvent (Fig. 3a and Fig. S7 in the ESI† for the fluorescence intensities and spectra, respectively). We confirmed that the photoirradiation induced a decrease of the absorbance of **1** (Fig. 3b and Fig.

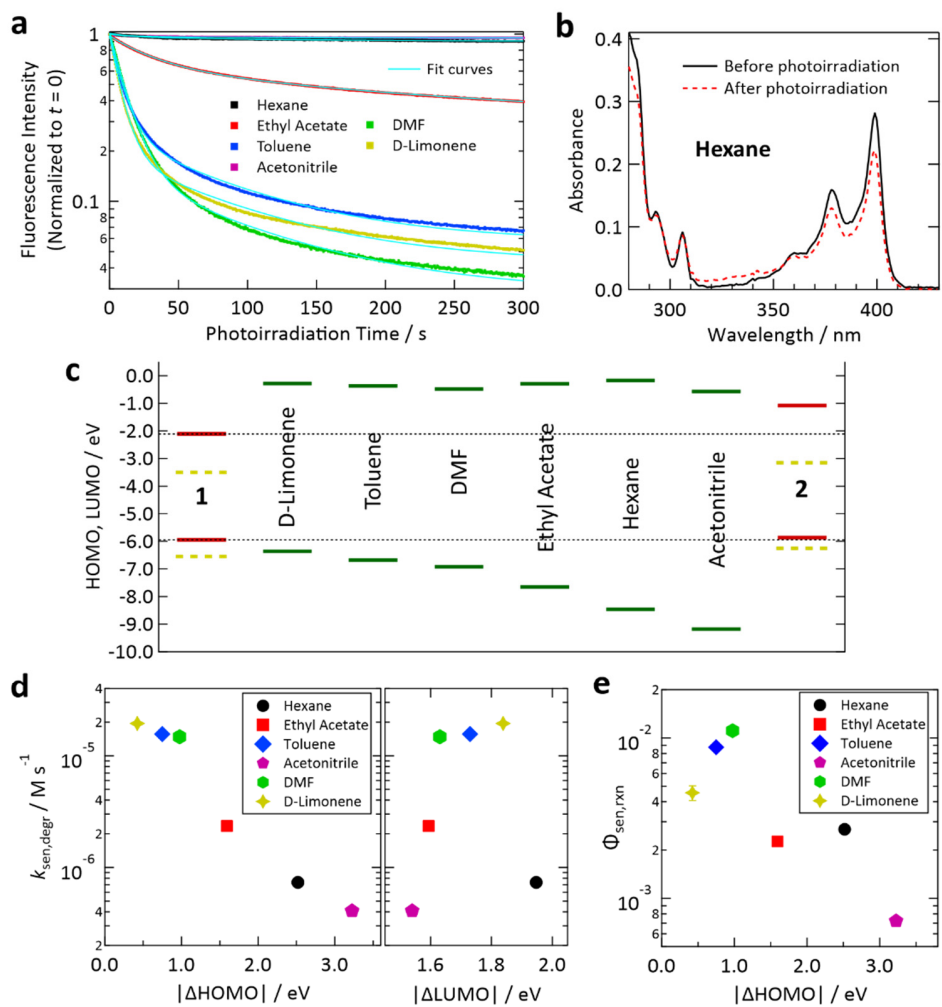

**Fig. 3** (a) Temporal decay profiles of the fluorescence intensity of deaerated samples containing only sensitizer **1** under continuous irradiation at 405 nm. The excitation light intensity for each case (cf. Table 1) was chosen so that the $T_1$ state of **1** was generated at a rate of $1.9 \times 10^{-3}$ M/s. (b) Decrease of the absorbance of **1** in deaerated hexane (optical path length: 1 mm) induced by photoirradiation at 405 nm using the setup and conditions described in Section 5 of the ESI†. (c) Calculated HOMO and LUMO levels of **1**, **2**, and the solvents. The SOMO levels of the $T_1$ states of **1** and **2** are shown as yellow dashed lines. (d) Degradation rates of fluorescence intensities determined from the results in panel (a) plotted against the difference between the HOMO levels of **1** and the solvent (left) and that between the LUMO levels of **1** and the solvent (right). (e) Reaction quantum yield between the $T_1$ state of **1** and solvent for the samples containing only **1** estimated using the optical absorption changes in Fig. S8 in the ESI†.

S8 in the ESI†), indicating that photoirradiation induced degradation of **1**. To compare the degradation rates of **1** in different solvents, we determined the rate of sensitizer degradation $k_{sen,degr}$ [M/s] from the decay curves in Fig. 3a by applying a double-exponential fit (see Section 11 of the ESI† for the procedure used to calculate $k_{sen,degr}$).

Here we use the frontier orbital theory to discuss the observed photoinduced degradation of **1** in the solvents. Generally, excited states of ketones such as the T$_1$ state of **1** have n,π* electronic configuration where n and π* are singly occupied molecular orbitals (SOMOs) and can serve as electron-accepting and -donating orbitals, respectively.[58] Generally, such SOMOs interact with the highest occupied molecular orbital (HOMO) and lowest unoccupied molecular orbital (LUMO) of an adjacent molecule and create new orbitals into which electrons from both molecules are partially or fully transferred; such a charge transfer generally allows energetic stabilization and may lead to formation of an excited-state complex.[58] For the n,π* state of ketones, such intermolecular interaction with a ground-state molecule such as a solvent molecule may cause hydrogen abstraction from the latter because of the half-filled orbital on the oxygen atom of the ketone. Hydrogen abstraction by ketones has been widely studied.[67–69] Two factors are known to govern this intermolecular reaction: (i) the energetic proximity of the frontier orbitals of the two interacting molecules and (ii) the constructive spatial overlap of these orbitals.[58]

To study factor (i), we calculated the HOMO and LUMO levels of **1**, **2**, and the solvents, as depicted in Fig. 3c. In this figure, SOMO levels of the T$_1$ states of **1** and **2** are also shown. From the relation between $k_{sen,degr}$ and the energetic separations of the HOMOs and LUMOs between **1** and the solvents (denoted as Δ|HOMO| and Δ|LUMO|, respectively), we found a clear correlation of $k_{sen,degr}$ with Δ|HOMO|, whereas no obvious correlation was found between $k_{sen,degr}$ and

Δ|LUMO| (Fig. 3d). The same tendency was also observed when the difference between the ionization energies of **1** and the solvents (which physically corresponds to Δ|HOMO|) and that between their electron affinities (which corresponds to Δ|LUMO|) were plotted (Fig. S9, ESI†). These results reveal that the electron transfer from the solvent to **1** is the rate-limiting step of this photodegradation, which can be interpreted as an electron transfer-initiated hydrogen abstraction process.[69,70] We also estimated the quantum efficiency of the degradation of the $T_1$ state of **1** in each solvent ($\Phi_{sen,rxn}$) from the decease of the absorbance of **1** induced by the controlled photoirradiation (cf. Fig. S4, ESI†). The procedure followed to calculate $\Phi_{sen,rxn}$ is described in Section 13 of the ESI†. Although the scatter of the data points is larger than that in the case of $k_{sen,degr}$, a similar correlation with Δ|HOMO| was also found for $\Phi_{sen,rxn}$ (Fig. 3e).

Next, we investigated photoinduced changes of samples containing both **1** and **2**. For the sample with hexane, photodegradation of **1** was suppressed by the presence of $2\times10^{-3}$ M of **2**, as recognized from the invariance of the optical absorption spectrum of **1** during photoirradiation (Fig. 4a). This suppression is ascribed to prompt TET from **1** to **2** in hexane, which drastically shortens the lifetime of the $T_1$ state of **1**, meaning that **2** strongly suppresses the probability of **1** reacting with the solvent. A similar tendency was also found for the samples with other solvents (Fig. S8 and S10 in the ESI†). However, for the sample with DMF, the decrease in the absorbance of **1** was not well suppressed (Fig. S10, ESI†); this could be because of inefficient TET from **1** to **2** caused by the relatively high polarity of DMF (cf. Fig. 2d and 1b). The suppressed photodegradation of **1** in hexane induced by addition of **2** was also evidenced by the invariance of the fluorescence emission intensity of **1** even after 80 min of photoirradiation (Fig. 4b); the similar tendency was also seen for the samples with other solvents (see Fig. S7 and S11 in the ESI†).

Our results reveal that by adding an energy-accepting emitter at sufficient concentration (of the order of 10$^{-3}$ M), preferable aspects of ketones as the sensitizer (cf. first paragraph of the Experimental section) can be harnessed for UV-UC while effectively suppressing the drawback of

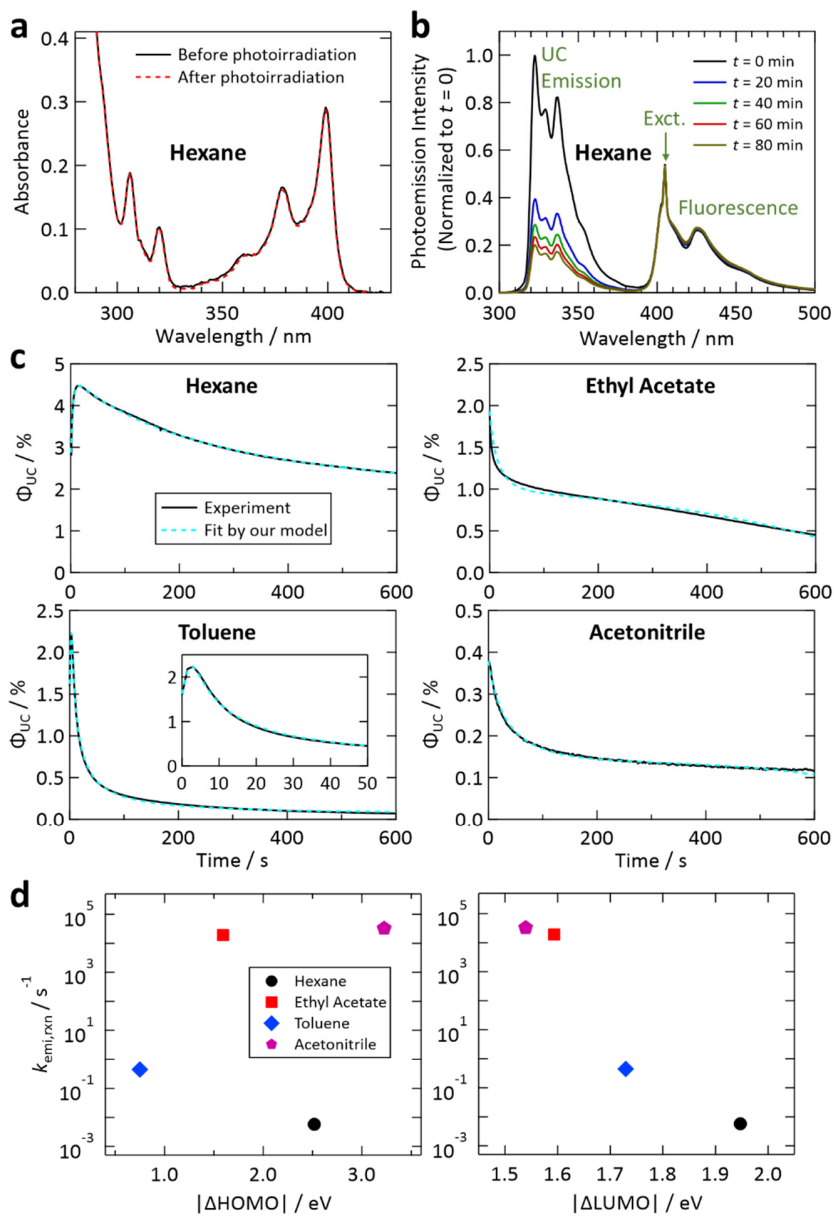

**Fig. 4** (a) Demonstration of the suppression of the decrease of the absorbance of **1** by addition of **2** in hexane using the same conditions as in Fig. 3b. (b) Demonstration of the suppression of the decrease of the fluorescence of **1** upon addition of **2** in hexane, which was measured using the same conditions as in Fig. 3a. See also Fig. S7 and S11 in the ESI†. (c) Fit of the temporal decay curves of the UC quantum efficiency shown in Fig. 2e (solid curves) by the kinetic model proposed in this study (dashed curves). (d) Reaction rates between the T$_1$ state of **2** and the solvents, obtained from the fittings shown in (c), plotted against the difference between the HOMO levels of **2** and the solvent (left) and that between the LUMO levels of **2** and the solvent (right).

using triplet ketones; i.e., the relatively high reactivity of their $T_1$ state. Considering the viscosities of the solvents employed in this study (ca. 0.3–0.6 mPa·s at room temperature), the diffusion-controlled rate constant $k_{\text{diff}}$ was estimated to be $1-2\times10^{10}$ $M^{-1}$ $s^{-1}$ using the following equation[16,58]

$$k_{\text{diff}} = \frac{8RT}{3000\eta} \qquad (1)$$

where $R$, $T$, and $\eta$ are the gas constant, temperature, and solvent viscosity, respectively. From the concentration of the energy acceptor **2** ($2\times10^{-3}$ M) and assuming Case A in Fig. 1b, the lifetime of the $T_1$ state of **1** was estimated to be only 25–50 ns, which supports the results in Fig. 4a and 4b.

The rate of the reaction between the $T_1$ state of **1** and ground state of **2** was considered to be negligible, even though their ground-state HOMO levels are close (Fig. 3c), for the following reasons. First, a bimolecular reaction rate is proportional to the product of the concentrations of the two species involved. In our samples, the concentration of **2** ($2\times10^{-3}$ M) was much lower than that of the solvents (7–20 M). Second, the interaction time between the $T_1$ state of **1** and ground state of **2** should be very short because such an encounter immediately causes an exothermic TET,[11] unlike the interaction between the $T_1$ state of **1** and the solvent, which can last much longer.

We have reached the point to discuss the temporal decay curves of the UC emission intensity under continuous photoirradiation in Fig. 2e. To analyze these decay curves, we developed the theoretical model described below. First, we postulate that the triplet emitter (E*) becomes a new species (ε) by reacting with a surrounding solvent molecule (sol) at a rate of $k_{\text{emi,rxn}}$ [$s^{-1}$]. This ε is assumed to quench both E* and the triplet sensitizer (S*) at the $k_{\text{diff}}$ given by eqn (1). Therefore,

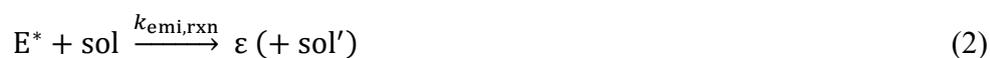

$$E^* + \text{sol} \xrightarrow{k_{\text{emi,rxn}}} \varepsilon\ (+\ \text{sol}') \qquad (2)$$

$$\text{E}^* + \varepsilon \xrightarrow{k_{\text{diff}}} \text{E} + \varepsilon^*; \text{S}^* + \varepsilon \xrightarrow{k_{\text{diff}}} \text{S} + \varepsilon^* \qquad (3)$$

Here, E and S are the ground states of the emitter and sensitizer, respectively, and $\varepsilon^*$ is the excited state of $\varepsilon$. In this model, E >> E* and S >> S* are assumed and the reaction between S* and solvent is neglected based on the considerations mentioned above. Photoirradiation of the sample was confirmed to shorten the triplet lifetime of **2** ($\tau_T$) (Section 16 of the ESI†). Furthermore, it is assumed that $\varepsilon^*$ converts into an inactive species ($\varepsilon_{\text{inactive}}$) at a quantum yield of $\Phi_{\varepsilon,\text{rxn}}$, presumably by reacting with the solvent as follows.

$$\varepsilon^* + \text{sol} \xrightarrow{\Phi_{\varepsilon,\text{rxn}}} \varepsilon_{\text{inactive}} \; (+ \text{sol}'') \qquad (4)$$

$$\varepsilon^* \xrightarrow{1-\Phi_{\varepsilon,\text{rxn}}} \varepsilon \qquad (5)$$

In addition, we consider initial impurity species in the solvent, Q, which quenches both E* and S*. Similar to the case of $\varepsilon$, we introduce the kinetic relations of

$$\text{E}^* + \text{Q} \xrightarrow{k_{\text{diff}}} \text{E} + \text{Q}^*; \text{S}^* + \text{Q} \xrightarrow{k_{\text{diff}}} \text{S} + \text{Q}^* \qquad (6)$$

$$\text{Q}^* + \text{sol} \xrightarrow{\Phi_{\text{Q},\text{rxn}}} \text{Q}_{\text{inactive}} \; (+ \text{sol}''') \qquad (7)$$

$$\text{Q}^* \xrightarrow{1-\Phi_{\text{Q},\text{rxn}}} \text{Q} \qquad (8)$$

We further assume that the second-order rate constant between E* molecules for the TTA process ($k_2$) is close to $k_{\text{diff}}$ (i.e., $k_2 \approx k_{\text{diff}}$), which was found to be a quantitatively good approximation.[36] Although the degradation phenomenon considered here is transient, the timescales of the above-described kinetics are much shorter than those of the change of the UC emission intensity. Therefore, at each instantaneous moment during continuous photoirradiation, the quasi-steady-state approximation is considered to hold well for E*,

$$\frac{d[\mathrm{E}^*]}{dt} \cong 0. \tag{9}$$

Combining all these relations, the proposed model describing the temporal change of UC emission intensity under continuous photoirradiation is obtained as

$$k_{\mathrm{diff}}[\mathrm{E}^*]^2 + \{k_{\mathrm{T}} + k_{\mathrm{diff}}([\varepsilon] + [\mathrm{Q}])\}[\mathrm{E}^*] - \left(\frac{[\mathrm{E}]}{[\mathrm{E}]+[\mathrm{Q}]+[\varepsilon]}\right)\Gamma = 0. \tag{10}$$

Here, $k_{\mathrm{T}}$ is the first-order decay rate of E* (= $\tau_{\mathrm{T}}^{-1}$), which was determined by time-resolved photoemission measurements using light pulses (cf. Experimental section). $\Gamma$ is the generation rate of the $T_1$ state of the sensitizer, which was $1.9 \times 10^{-3}$ M/s. Eqn (10) is a quadratic equation of [E*] and thus [E*] at each moment could be expressed using the other parameters in this equation. These parameters were numerically calculated at various times ($t$) after the onset of the photoirradiation at $t = 0$ (see Section 17 of the ESI† for details of the calculation). Because UC emission intensity is proportional to [E*]$^2$, the theoretical curve of the UC emission intensity for $t > 0$ can be obtained. Finally, the theoretical curve was computationally fitted to the experimental curve by treating $k_{\mathrm{emi,rxn}}$, $\Phi_{\varepsilon,\mathrm{rxn}}$, $\Phi_{\mathrm{Q,rxn}}$, and Q as adjustable parameters.

Figure 4c shows the results of the fittings of this model to the experimental curves for the samples with hexane, ethyl acetate, toluene, and acetonitrile. In all cases, the agreement between the model and experimental curves was good, suggesting that the model has captured the physicochemical characteristics of the present system. All these fittings resulted in Q ≤ $5 \times 10^{-4}$ M (i.e., molar fraction of 0.005% or lower), which does not contradict the certified purities of the solvents. It is noted that processes represented by eqn (4) and (7) are necessary to fit our theoretical model to the experimental curves.

In Fig. 4d, the values of $k_{emi,rxn}$ obtained from the fitting are plotted against |ΔHOMO| and |ΔLUMO|; a correlation was found only for |ΔLUMO| and $k_{emi,rxn}$. The same tendency was also observed when the difference between the ionization energies of **2** and the solvents and that between the electron affinities of **2** and the solvents were plotted (Fig. S13, ESI†). These results suggest that the process described by eqn (2) is limited by electron transfer from **2** to the solvent; i.e., electron transfer in the opposite direction to that in the reaction between **1** and the solvent discussed above. We did not carry out further detailed investigation of the reaction mechanism because it is beyond the scope of the present study. Nevertheless, the findings acquired from our experimental and theoretical investigations revealed that the photostability of this UV-UC system is controlled by the energetic difference between the relevant frontier orbital levels of the solute (**1** or **2**) and solvent, and that these photodegradation reactions are rate-limited by the electron transfer between molecules.

## 4. Conclusions

Using sensitizer **1** and emitter **2**, UV-UC to a shorter wavelength than 340 nm (maximum intensity at 322 nm) was achieved in various solvents. Both $\Phi_{UC}$ and the photostability of **1** under continuous photoirradiation depended on the solvent. The use of hexane yielded the highest $\Phi_{UC}$ of 8.2%, which is close to that of 10.2% reported for UV-UC in the 350–400 nm range achieved using a nanocrystal sensitizer and PPO,[51] and also the highest photostability among the tested solvents. We found that $\Phi_{UC}$ was mainly governed by solvent polarity, which varied the relative $T_1$ energy-level matching between **1** and **2** because of the solvatochromic shift imposed on **1**. The solvent dependence of $\Phi_{T,sen}$ and $\Phi_{F,emi}$ should also affect $\Phi_{UC}$.

When the samples were prepared without **2**, $k_{\text{sen,degr}}$ was large in most of the tested solvents and found to be correlated with $\Delta|\text{HOMO}|$ between **1** and the solvent. This correlation indicated that the photodegradation of **1** was rate-limited by electron transfer from the solvent to **1** and likely to be an electron transfer-initiated hydrogen abstraction process. However, when the energy acceptor **2**, which quenches the $T_1$ state of **1**, was added to the samples, the degradation of **1** was effectively suppressed. This finding justifies the use of a ketonic sensitizer for UV-UC as long as the emitter concentration is higher than the order of $10^{-3}$ M in non-viscous solvents.

We developed a theoretical model and the curves generated by this model fitted the experimentally acquired temporal decay curves of the UC emission intensity well. This fitting provided several insights into the characteristics of the present UV-UC system. For example, the initial rapid rise of the UC emission intensity for the sample with hexane (cf. Fig. 4c) was ascribed to the presence of a trace amount of impurities (Q ~$1.9 \times 10^{-7}$ M). Furthermore, $k_{\text{emi,rxn}}$ obtained from the fitting was correlated with $\Delta|\text{LUMO}|$, which revealed that the photodegradation of **2** was rate-limited by electron transfer to the solvent. These findings indicate that the energetic difference between the frontier orbitals of the solute and solvent is the primary factor determining the photostability. Besides this viewpoint, the frontier orbital theory also addresses the importance of spatial overlap between two frontier orbitals involved in a reaction. Decreasing such overlap by addition of bulky groups to solutes may enhance their photostability.

Overall, this experimental and theoretical study has provided several fundamental insights regarding UV-UC in solvents. As general design guidelines for sample development, one should optimize solvent polarity to maximize $\Phi_{\text{UC}}$ and use a combination of solvent and solute whose frontier energy levels are as far apart as possible to enhance solute photostability. These guidelines have not previously been explicitly proposed for UV-UC or visible-to-visible UC. The physicochemical insights obtained from this study will help to establish stable and efficient UV-UC systems in the future.

## Conflicts of interest

There are no conflicts to declare.

## Acknowledgements

We cordially thank Prof. Susumu Kawauchi (Tokyo Institute of Technology) for valuable comments and discussion regarding the quantum-chemical simulations. We also thank Dr. Natasha Lundin for correcting the English used in this report.

Electronic Supplementary Information

# Visible-to-ultraviolet (<340 nm) photon upconversion by triplet–triplet annihilation in solvents

Yoichi Murakami,[1,]* Ayumu Motooka,[1] Riku Enomoto,[1] Kazuki Niimi,[2] Atsushi Kaiho[2] and Noriko Kiyoyanagi[2]

[1] School of Engineering, Tokyo Institute of Technology, 2-12-1 Ookayama, Meguro-ku, Tokyo 152-8552, Japan.

[2] Nippon Kayaku Co., Ltd., 3-31-12 Shimo, Kita-ku, Tokyo 115-8588, Japan.

*Corresponding Author: Yoichi Murakami, E-mail: murakami.y.af@m.titech.ac.jp

**List of Contents**

1. Photostability of visible-to-visible UC in an ionic liquid (Fig. S1)
2. Photostability of UV-UC using biacetyl and PPO in DMF (Fig. S2)
3. Optical absorption spectra of the sensitizer and emitter used in this study (Fig. S3)
4. Information about the solvents used in this study (Table S1)
5. Experimental setup to controllably induce photodegradation (Fig. S4)
6. Calculated dipole moments of the sensitizer and emitter (Fig. S5 and Table S2)
7. Determination of $\Phi_{UC}$

**1. Photostability of visible-to-visible UC in an ionic liquid**

We have reported several examples of visible-to-visible photon upconversion (UC) by triplet–triplet annihilation (TTA) using ionic liquids as the solvent.[S1–S4] To underpin the motivation of the present study, this supplementary section considers the photostability of such visible-to-visible UC and then the contrasting low photostability of visible-to-ultraviolet UC (UV-UC).

The inset of Fig. S1b shows a photograph of the sample used here, which was prepared and sealed in a quartz tube with a 2×2 mm square cross section on October 30, 2012, according to the procedure described prevously.[S1–S4] This sample was prepared using *meso*-tetraphenyltetrabenzoporphyrin palladium (PdPh$_4$TBP) as the sensitizer and perylene as the emitter with concentrations of 5×10$^{-5}$ M and 3×10$^{-3}$ M, respectively, dissolved in the ionic liquid 1-butyl-2,3-dimethylimidazolium bis(trifluoromethylsulfonyl)amide ([C$_4$dmim][NTf$_2$]). The molecular structures of these materials are shown in Fig. S1a. Since October 2012, this sample has been located on a desktop in our laboratory, where it is exposed to light from fluorescence tubes and weak indirect natural sunlight from outside. This sample still displays similar UC behavior to that at the time of the preparation. The photograph in the inset of Fig. S1b was taken on September 8, 2020, showing the sample was upconverting incident red light (633 nm, ca. 5 mW) into blue emission (around 450–480 nm), demonstrating its long lifetime (>7 years).

Figure S1b illustrates the excellent stability of the UC emission from this sample under continuous irradiation of laser light at 633 nm (3 mW; intensity: ca. 0.6 W/cm$^2$). This experiment was carried out on February 22, 2020. In contrast, when the sensitizer **1** and emitter **2** used in this article were dissolved in the same ionic liquid at concentrations of 2×10$^{-4}$ and 2×10$^{-3}$ M, respectively (see the Experimental section in the main text), the photostability of the sample was low, as shown by the purple curve in Fig. S1b, which is the same curve as that shown in Fig. 2e of the main text.

This comparison reveals that photodegradation in TTA-UC samples primarily depends on the characteristics of solutes, where molecules used for UV-UC have higher triplet energies than those used for visible-to-visible UC. Furthermore, in the UV-UC explored here, we found that the photostability and UC quantum efficiency ($\Phi_{UC}$) strongly depended on the solvent, which is the subject of the present study.

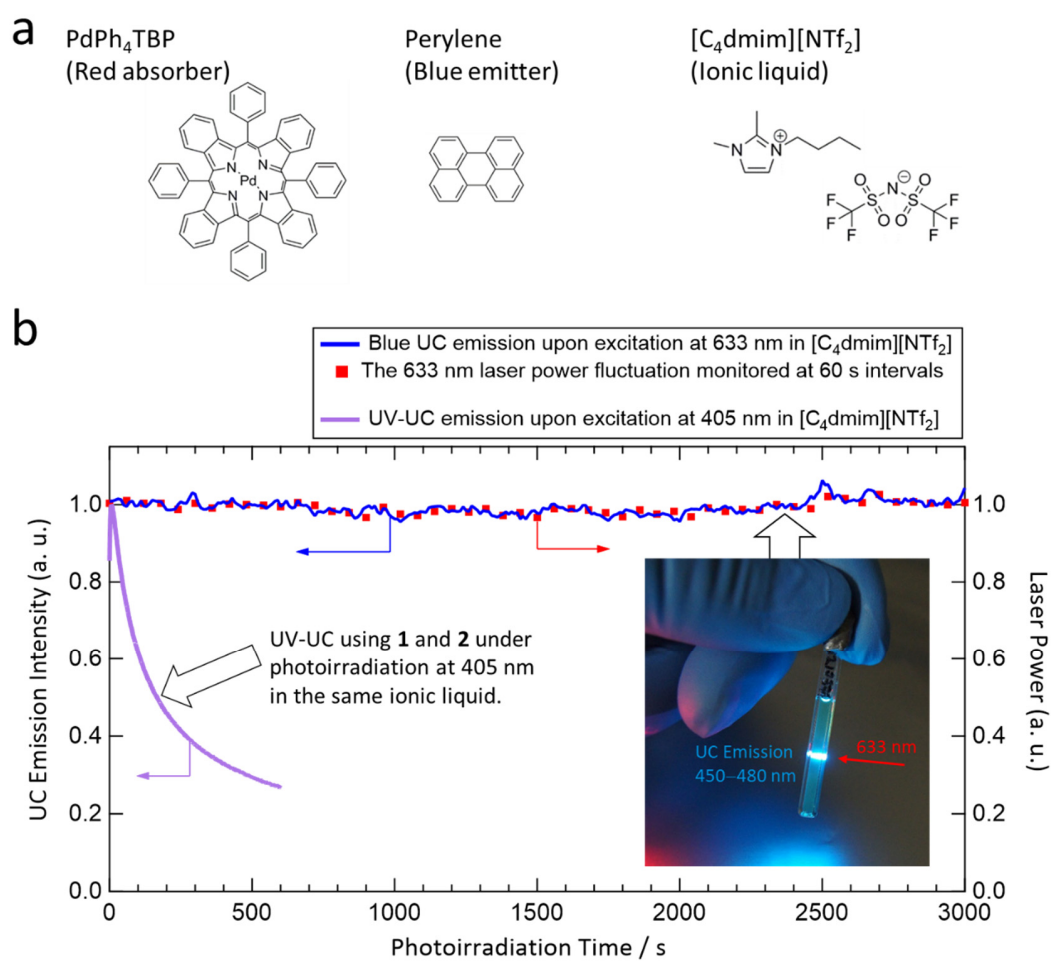

**Figure S1.** (a) Molecular structures of the sensitizer, emitter, and ionic liquid used here. (b) Temporal UC emission intensity profile acquired from the sample under continuous photoirradiation at 633 nm (blue curve; laser power: 3 mW) along with the simultaneously monitored temporal fluctuation of the laser power (red dots). For comparison, the temporal profile of the UV-UC emission intensity from a sample prepared using the sensitizer **1** and emitter **2** (see the main text for details) in the same ionic liquid under continuous 405-nm irradiation is also shown (purple curve; laser power: 2.2 mW), which is the same curve as that shown in Fig. 2e of the main text. Inset is a photograph of the ionic liquid sample measured here.

**2. Photostability of UV-UC using biacetyl and PPO in DMF**

To date, several examples of UV-UC using 2,5-diphenyloxazole (PPO), which generates UV emission around 350–400 nm, as the emitter have been reported.[S5–S10] The most representative sensitizer combined with PPO is 2,3-butanedione (biacetyl), as used in the pioneering work by Singh-Rachford and Castellano.[S5] In ref. S5, the authors used benzene as the solvent, presumably to decrease the rate of hydrogen abstraction by the triplet solutes, and reported $\Phi_{UC}$ of 0.58%. However, benzene is inappropriate for applications. The other reports that combined biacetyl and PPO used dimethylformamide (DMF) as the solvent.[S7,S9] To support our statements in the Introduction section of the main text, here we present our results for UV-UC samples prepared using biacetyl and PPO in DMF. All the samples used here were deaerated by nine freeze-pump-thaw (FPT) cycles by the method described in the Experimental section of the main text and measured using the same conditions as those used for other samples investigated in this report.

First, we investigated the sample containing only biacetyl at a concentration of $2\times10^{-3}$ M (Fig. S2a–c). Using the setup described in Section 5 of this Supplementary Information, an expanded 405-nm laser beam was irradiated onto the sample under the conditions described therein, which were the same as those used in Fig. 3b and 4a of the main text. After this photoirradiation, the absorbance of biacetyl had disappeared (Fig. S2a). We also measured the temporal changes of the fluorescence spectrum and intensity (Fig. S2b and S2c, respectively) for this sample sealed in a 1×1-mm glass capillary exposed to an excitation power at 405 nm that induced a triplet generation rate of biacetyl of ca. $1.65\times10^{-3}$ M/s (i.e., slightly weaker excitation conditions than those used for Fig. 3a in the main text and Fig. S7 below). The fluorescence quickly diminished during the photoirradiation. These results indicate the low photostability of biacetyl in DMF.

Next, we investigated a sample containing both biacetyl and PPO with concentrations of $2\times10^{-3}$ and $8\times10^{-3}$ M, respectively, which are the same concentrations as those used in ref. S5. The absorption spectrum of this sample is shown in Fig. S2d. Under the same photoirradiation conditions (i.e., triplet generation rate of ca. $1.65\times10^{-3}$ M/s on biacetyl), the UC emission rapidly diminished, almost disappearing within 30 s.

Overall, based on the results presented here, photodegradation in UV-UC is an important issue to investigate and understand. Thus, the issue of photodegradation is not limited to the particular case of **1** and **2** employed in this study. Recently, Lee et al.[S11] also presented a report on this aspect of UV-UC.

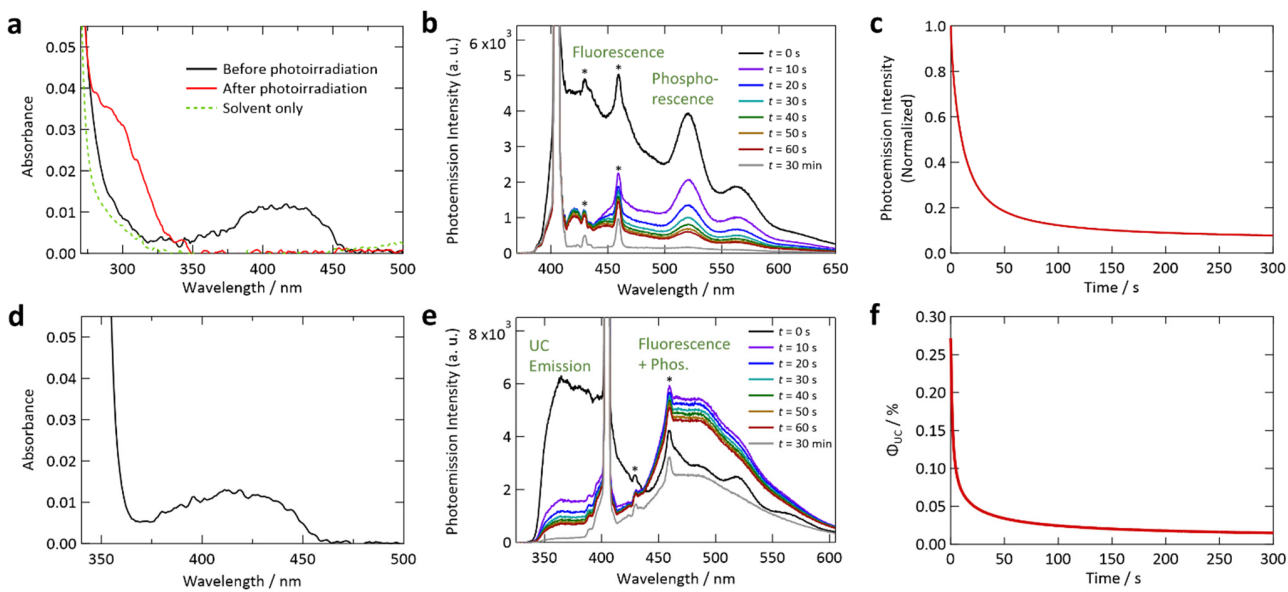

**Figure S2.** (a) Change of the optical absorption of a sample containing only biacetyl ($2\times10^{-3}$ M) in deaerated DMF induced by photoirradiation at 405 nm. (b) Temporal change of the fluorescence spectrum of this sample sealed in a 1×1-mm glass capillary under continuous irradiation at 405 nm and (c) temporal profile of the fluorescence intensity spectrally integrated between 475 and 625 nm. (d) Optical absorption spectrum of a sample containing both biacetyl ($2\times10^{-3}$ M) and PPO ($8\times10^{-3}$ M) in deaerated DMF. (e) Temporal change of the photoemission spectrum of the sample sealed in a 1×1-mm glass capillary under continuous irradiation at 405 nm and (f) temporal profile of the UC quantum efficiency. In (a) and (d), the optical path length was 1 mm. In (b) and (e), the sharp peaks marked with asterisks were unidentified and may be either Raman scattering from the sample or sidebands from the laser light source.

### 3. Optical absorption spectra of the sensitizer and emitter used in this study

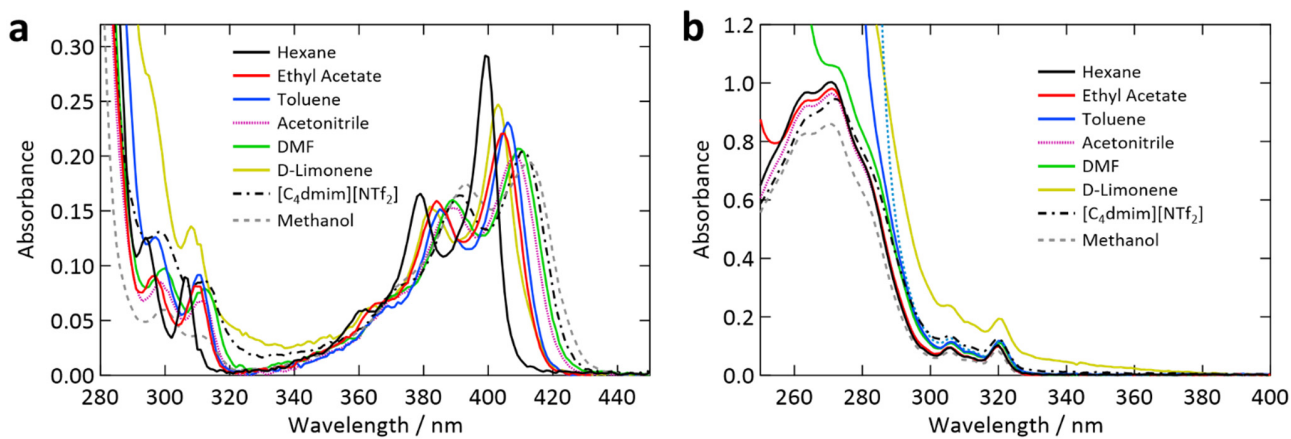

**Figure S3.** (a) Optical absorption spectra of the sensitizer **1** and (b) emitter **2** measured in different solvents at concentrations of $2\times10^{-4}$ and $2\times10^{-3}$ M, respectively. The optical path length was 1 mm.

### 4. Information about the solvents used in this study

Information about the solvents used in this report is summarized in Table S1. The refractive index values were used to calculate $\Phi_{UC}$ in Section 7 of this Supplementary Information.

**Table S1.** Information about the solvents used in this study

| Solvent | Supplier | Purity (Supplier product #) | Refractive index |
|---|---|---|---|
| Hexane | Supelco | ≥ 99.7 % (GC) (52750) | |
| Hexane (reference used in Fig. S6) | TCI | > 99.5 % (GC) (S0278) | $1.373^a$ |
| Hexane (reference used in Fig. S6) | Sigma-Aldrich | ≥ 95.0 % (GC) (13-0800-5) | |
| Ethyl Acetate | Wako | 99.8+ % (GC) (055-05991) | $1.372^a$ |
| Toluene | Wako | 99.8+ % (GC) (208-12871) | $1.497^a$ |
| Acetonitrile | Wako | 99.8+ % (GC) (018-22901) | 1.339[a] |
| Dimethylformamide (DMF) | Sigma-Aldrich | ≥ 99.90 % (GC) (270547) | 1.421[a] |
| D-Limonene | TCI | > 99.0 % (GC) (L0105) | 1.474[a] |
| [C$_4$dmim][NTf$_2$] | Merck | ≥ 98.0 % (HPLC) (490288) | 1.435[b] |
| Methanol | Wako | 99.9+ % (GC) (139-13995) | 1.329[a] |

[a] From the PubChem website (URL: https://pubchem.ncbi.nlm.nih.gov). All data were collected at the sodium D-line. Temperatures for these values were 30 °C (acetonitrile), 25 °C (D-limonene), 20 °C (ethyl acetate), 25 °C (hexane), 18 °C (mesitylene), 20 °C (methanol), 25 °C (DMF), and 20 °C (toluene).
[b] From ref. S1; at the sodium D-line at 21 °C.

## 5. Experimental setup to controllably induce photodegradation

Figure S4 illustrates the setup used to controllably induce photodegradation by irradiating an expanded 405-nm laser beam onto almost the entire volume of the sample liquid (2 mL) in a hermetically sealed glass vial (capacity: 6 mL). The liquid height in the vial was ca. 10 mm. The liquid sample was deaerated by conducting FPT cycles just before it was transferred into the hermetically sealed vial; this transfer was promptly carried out inside the vacuum-type SUS glovebox filled with fresh nitrogen gas (see the Experimental section of the main text). As illustrated, the expanded light beam (diameter: ca. 5 mm; power: ca. 22 mW) was incident from the bottom of the vial so that the light was entirely absorbed by the sample. The photoirradiation was continued until each molecule of sensitizer **1** converted to the triplet state 85 times on average. The duration of photoirradiation was chosen assuming that the initial absorbance of **1** at 405 nm did not change during the course of irradiation.

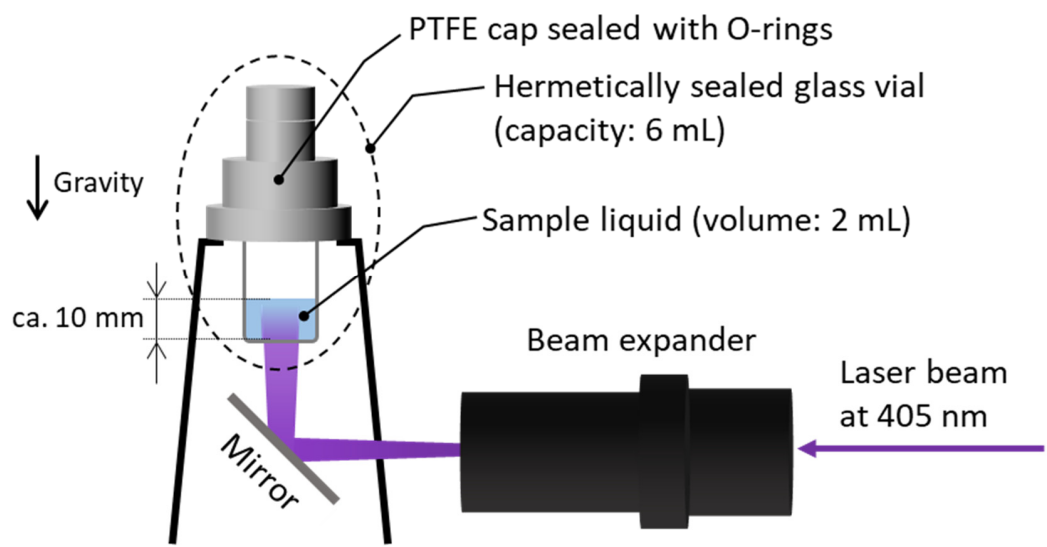

**Figure S4.** Schematic illustration of the setup to controllably induce photodegradation of a sample liquid.

## 6. Calculated dipole moments of the sensitizer and emitter

Dipole moments of the sensitizer **1** and emitter **2** were calculated using Gaussian 16® at the B3LYP/6-31G++(d,p) level, as summarized in Table S2. The corresponding graphics are shown in Fig. S5, where the blue arrows represent dipole moment vectors.

**Table S2.** Calculated dipole moments for the sensitizer **1** and emitter **2**

|  | Electronic State | Dipole Moment (Debye) |
|---|---|---|
| Sensitizer **1** | $S_0$ | 7.166 |
|  | $S_1$ | 7.769 |
|  | $T_1$ | 8.348 |
| Emitter **2** | $S_0$ | 0 |
|  | $S_1$ | 0 |
|  | $T_1$ | 0 |

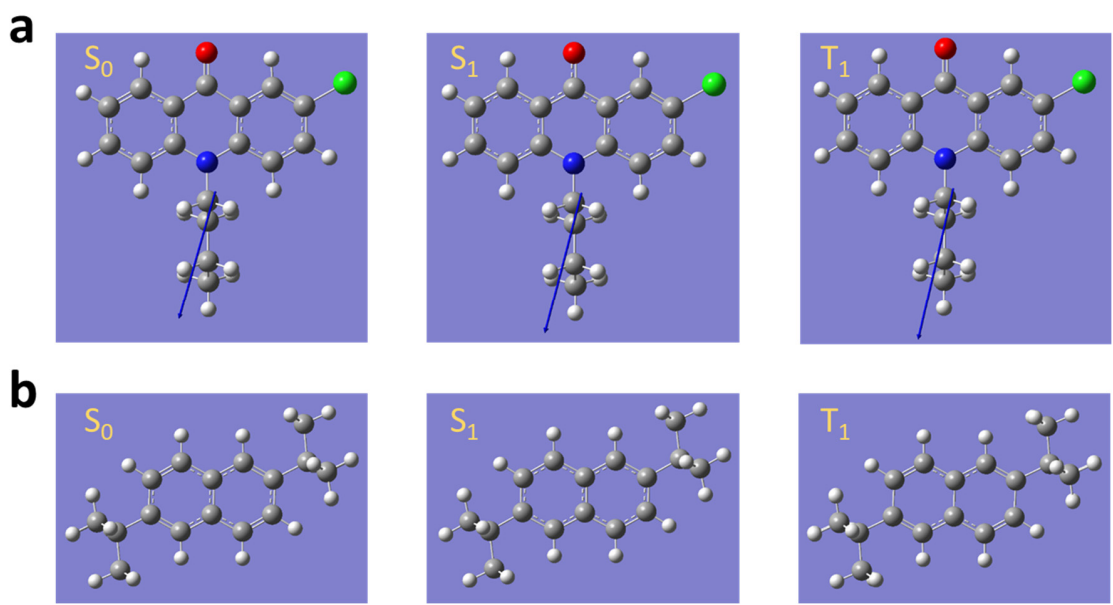

**Figure S5.** Optimized molecular structures and dipole moments (blue arrows) for (a) sensitizer **1** and (b) emitter **2**.

## 7. Determination of $\Phi_{UC}$

The upconversion quantum efficiency $\Phi_{UC}$ (with a defined maximum of 100%) in this article was determined using the following standard relationship.[S12]

$$\Phi_{UC} = 2\Phi_R \left(\frac{1-10^{-A_R}}{1-10^{-A_{UC}}}\right)\left(\frac{I_{UC}^{Em}}{I_R^{Em}}\right)\left(\frac{I_R^{Ex}}{I_{UC}^{Ex}}\right)\left(\frac{h\nu_{UC}}{h\nu_R}\right)\left(\frac{n_{UC}}{n_R}\right)^2 \quad (S1)$$

Here, $\Phi_R$, $A$, $I^{Em}$, $I^{Ex}$, $h\nu$, and $n$ represent the fluorescence quantum yield of a reference sample, absorbance, photoemission intensity, excitation light intensity, photon energy at the excitation wavelength, and the refractive index of the solvent, respectively. The subscripts "UC" and "R" represent an UC sample and reference, respectively. For the second term on the right-hand side, we used $1-10^{-A}$, which is absorptance, instead of its mathematically approximated form of $A$ (see ref. S12 for further details).

We used a toluene solution of 9,10-diphenylanthracene (concentration: $4\times10^{-4}$ M) deaerated by FPT cycles as the reference sample, which was determined to have $\Phi_R$ of 0.940 at the excitation wavelength of 405 nm using our absolute quantum yield spectrometer (Quantaurus-QY, Hamamatsu). The values of $n$ were taken from Table S1. The emission intensity between 310 and 380 nm was used to calculate $\Phi_{UC}$; i.e., the emission between 380 and 405 nm was not used to exclude the tail of the fluorescence and thermally induced UC emission. All photoemission spectra in this report, including those used to determine $\Phi_{UC}$, were corrected by the wavelength-dependent sensitivities of the grating in our monochromator and CCD array detector as reported previously.[S1–S4]

## 8. Effect of solvent purity on temporal decay profiles of $\Phi_{UC}$

Figure S6 compares temporal decay profiles of $\Phi_{UC}$ acquired from three samples prepared under the same conditions using hexane of different purity grades (cf. Table S1). The black curve is the same as that shown in Fig. 2e of the main text. The results reveal that the solvent purity affected the magnitude of $\Phi_{UC}$, especially when low-purity hexane ($\geq$ 95%, in green) was used, but it did not change the qualitative character of the temporal decay profile.

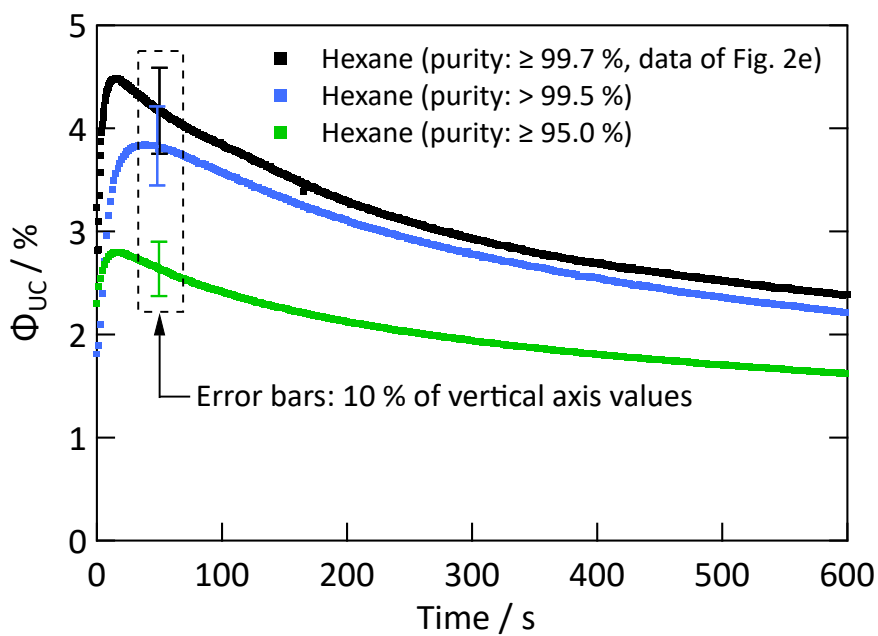

**Figure S6.** Effect of solvent purity on the decay profiles of $\Phi_{UC}$ measured for three samples prepared using hexane with different purity grades (cf. Table S1). The black curve is the data presented in Fig. 2e of the main text.

## 9. Temporal changes of fluorescence spectra of the sensitizer in the absence of the emitter during photoirradiation in different solvents

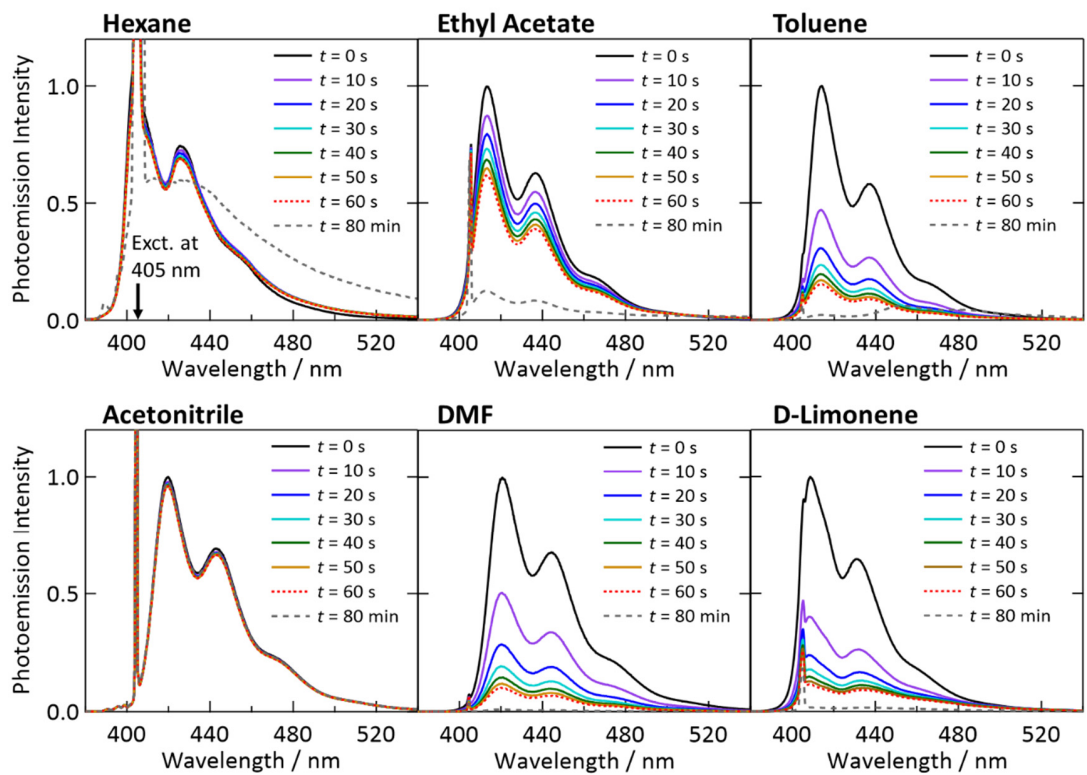

**Figure S7.** Temporal decay of the fluorescence spectra of sensitizer **1** acquired under continuous irradiation at 405 nm of samples without emitter **2** sealed in glass capillaries. These results were used to generate the temporal decay curves in Fig. 3a of the main text.

## 10. Photoirradiation-induced changes of optical absorption spectra of samples containing only the sensitizer in different solvents

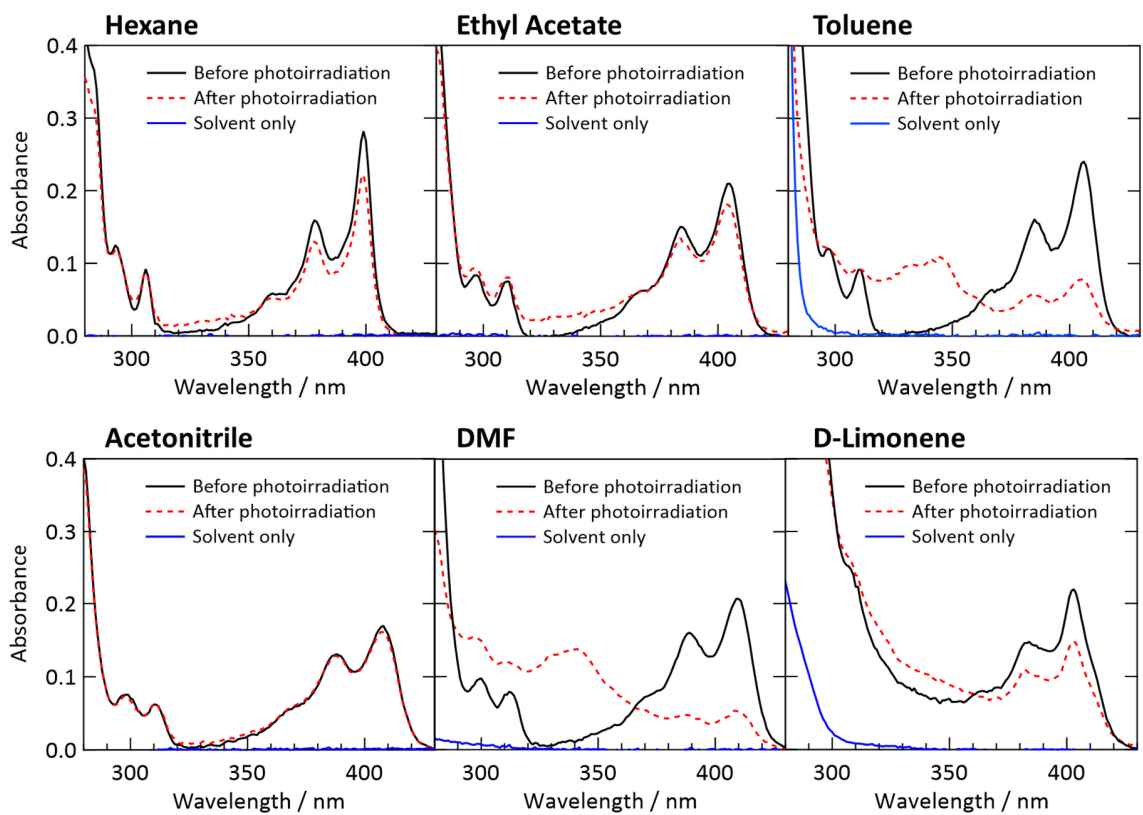

**Figure S8.** Comparison of optical absorption spectra of samples containing only sensitizer **1** before and after irradiation with 405-nm light. These experiments were carried out using the experimental setup and conditions described in Section 5 of this Supplementary Information. The result for hexane is the same as that shown in Fig. 3b of the main text.

## 11. Procedure to calculate $k_{\text{sen,degr}}$

Here we describe the procedure used to calculate the photodegradation rate of sensitizer **1** during irradiation with 405-nm laser light from the fluorescence intensity decay curves shown in Fig. 3a of the main text. As mentioned in the main text, these curves were acquired under the same excitation condition; that is, the triplet state of **1** was generated at a rate of ca. $1.9 \times 10^{-3}$ M/s. Our aim here is to estimate the consumption rate of the sensitizer molecules under this excitation condition, which is denoted as $k_{\text{sen,degr}}$ [mol/(L·s) = M/s]. The consumption of **1** is ascribed to the chemical reaction between **1** in the triplet state and the solvent, as discussed in the main text.

To estimate $k_{\text{sen,degr}}$, we fitted the normalized experimental fluorescence intensity decay curves shown in Fig. 3a of the main text with the following double-exponential function

$$I(t) = y_0 + A_1 \exp(-k_1 t) + A_2 \exp(-k_2 t). \qquad (S1)$$

Although the real photophysics should be described by more complex kinetic equations, as discussed in the main text, the present procedure is sufficient to obtain values of $k_{\text{sen,deg}}$. As illustrated by the fitting curves in Fig. 3a of the main text, eqn (S1) fitted the experimental fluorescence decay curves well in all cases. In eqn (S1), the relation $y_0 + A_1 + A_2 = 1$ holds by definition and the initial condition $I(0) = 1$ corresponds to the initial sensitizer concentration of $2\times10^{-4}$ M.

Then, we employed two reasonable assumptions that (i) the intensity of the fluorescence, which arose from the $S_1$ state, was proportional to the concentration of intact **1** in the solution, and thus that (ii) both constants $k_1$ and $k_2$, although phenomenological, provide quantitative information about the consumption rate of intact **1**. Based on these assumptions, the degradation rate of **1** at $t = 0$ (i.e., when the sensitizer concentration was $2\times10^{-4}$ M), $k_{\text{sen,degr}}$, was calculated from the relation

$$k_{\text{sen,degr}} = C_0 \times (A_1 k_1 + A_2 k_2). \qquad (S2)$$

$$\left[\frac{\text{mol}}{\text{L}\cdot\text{s}}\right] \quad \left[\frac{\text{mol}}{\text{L}}\right] \quad \left[\frac{1}{\text{s}}\right]$$

Here, $C_0$ is the initial sensitizer concentration of $2\times10^{-4}$ M. Table S3 summarizes the fitting results and calculated values of $k_{\text{sen,degr}}$ for **1** in different solvents.

Table S3. Results of fittings by eqn (S1) and $k_{sen,degr}$ calculated from eqn (S2) for **1** in different solvents.

| Solvent | $A_1$ | $k_1$ / s | $A_2$ | $k_2$ / s | $k_{sen,degr}$ / M s$^{-1}$ |
|---|---|---|---|---|---|
| Hexane | 0.06535 | 0.007170 | 0.04374 | 0.07327 | $7.347 \times 10^{-7}$ |
| Ethyl Acetate | 0.3550 | 0.003677 | 0.3449 | 0.03024 | $2.347 \times 10^{-6}$ |
| Toluene | 0.1976 | 0.01187 | 0.7112 | 0.1066 | $1.563 \times 10^{-5}$ |
| Acetonitrile | 0.02419 | 0.01081 | 0.0234 | 0.07627 | $4.092 \times 10^{-7}$ |
| Dimethylformamide (DMF) | 0.1262 | 0.01077 | 0.8627 | 0.08417 | $1.479 \times 10^{-5}$ |
| D-Limonene | 0.1467 | 0.01125 | 0.7895 | 0.1212 | $1.947 \times 10^{-5}$ |

## 12. Plots of $k_{sen,degr}$ against ionization energy and electron affinity

The results in Fig. 3d of the main text were presented based on HOMO and LUMO levels. Although the representation using HOMOs and LUMOs is easy to understand intuitively, in general, the quantitative reliability of orbital energy levels is affected by the choice of the basis set and level of theory used in the calculation. (In this report, all quantum-chemical calculations were performed using Gaussian 16® at the B3LYP/6-31G++(d,p) level.)

To alleviate this concern, use of the ionization energy (IE) and electron affinity (EA), which physically correspond to HOMO and LUMO energies, respectively, can enhance the quantitative reliability of analysis. This is because both IE and EA are calculated based on the total energy of the molecule considered, which means they are less affected by the choice of the basis set and calculation level than calculated HOMO and LUMO energies. Specifically, IE can be calculated by subtracting the energy of the neutral ground-state species from that of the radical cation species, and EA can be calculated by subtracting the energy of the radial anion species from that of the neutral ground-state species. Here, energies of the radial cation and radical anion were calculated using the molecular structure of the neutral ground-state species (i.e., vertical assumption).

Figure S9 shows plots of $k_{sen,degr}$ against the difference between the IEs (left, corresponding to Δ|HOMO|) of **1** and the solvents and that between the EAs (right, corresponding to Δ|LUMO|) of **1** and the solvents. We observed that $k_{sen,degr}$ was correlated with the difference of IEs, whereas no correlation of $k_{sen,degr}$ with the difference of EAs was found, supporting the results in Fig. 3d of the main text.

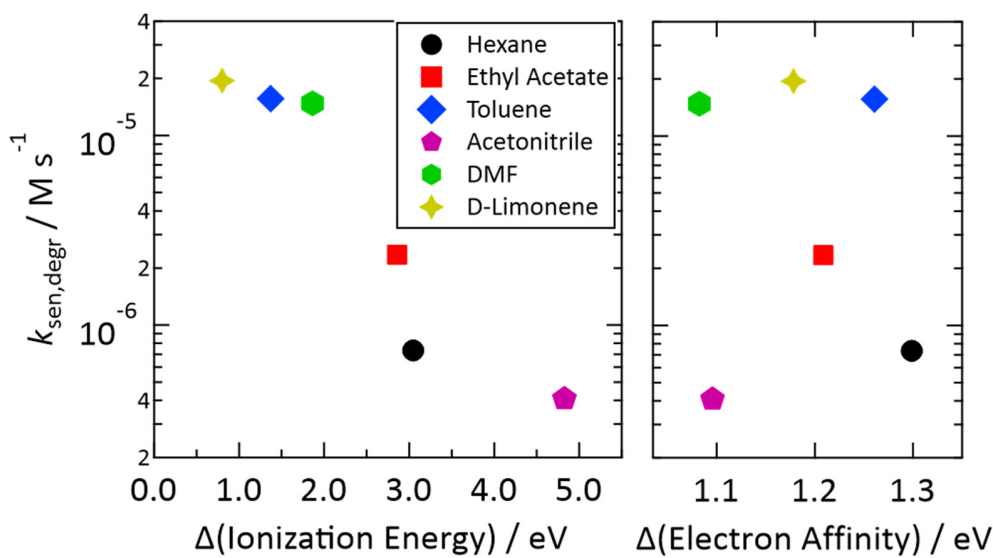

**Figure S9.** Degradation rates of the fluorescence intensities determined from the results in Fig. 3a of the main text, plotted against the difference between the ionization energies of **1** and the solvents (left) and the difference between the electron affinities of **1** and the solvents (right). See also Fig. 3d in the main text.

### 13. Procedure to calculate $\Phi_{sen,rxn}$

The experiments in Fig. S8 above were carried out by the method described in Section 5 of this Supplementary Information. As written therein, the photoirradiation time for each experiment was chosen assuming that the absorbance of **1** at 405 nm did not change during photoirradiation. To estimate the reaction quantum yield of the $T_1$ state of **1** and solvent ($\Phi_{\text{sen,rxn}}$) from the results of Fig. S8, the effect of using this assumption needs to be corrected. The details of this procedure are presented below.

First, we introduce the molar quantity of the intact sensitizer in the test vial of Fig. S4, denoted as $z$, which is a function of time $t$ and thus $z(t)$. The initial value $z(0)$ is $(2\times10^{-4}$ mol/L$)\times(2\times10^{-3}$ L$) = 4\times10^{-7}$ mol. We also introduce the absorbance of the sample liquid with an optical path length of 10 mm (cf. Fig. S4) at a wavelength of 405 nm, denoted as $A$, which is also a function of time and thus $A(t)$. The initial value $A(0)$ was calculated from $A_{\text{405nm}}$ in Table 1 of the main text. Using these parameters, $z(t)$ and $\Phi_{\text{sen,rxn}}$ were related with each other by

$$N_A \frac{dz}{dt} = -G_{\text{ph}}(1 - 10^{-A})\Phi_{\text{T,sen}}\Phi_{\text{sen,rxn}}. \tag{S3}$$

Here, $N_A$ is the Avogadro constant, $G_{\text{ph}}$ is the number of photons at 405 nm incident to the sample per unit time, and $\Phi_{\text{T,sen}}$ is the triplet quantum yield of **1** listed in Table 1 of the main text. Furthermore, there is a relationship of

$$A = \sigma z \Leftrightarrow z = \frac{A}{\sigma}, \tag{S4}$$

where $\sigma$ is a proportionality constant with a unit of mol$^{-1}$. $\sigma$ depends on the solvent and was in the range of ca. $1.6-6\times10^{6}$ mol$^{-1}$ in the present study. By substituting eqn (S4) into eqn (S3), we obtain

$$\frac{dA}{dt} = -\gamma(1 - 10^{-A}) \tag{S5}$$

where

$$\gamma = \frac{G_{ph}\Phi_{T,sen}\sigma}{N_A}\Phi_{sen,rxn}. \qquad (S6)$$

On the right-hand side of eqn (S6), all parameters except $\Phi_{sen,rxn}$ are known. Thus, the parameter $\gamma$ in eqn (S5) is an undetermined constant that is the function of only $\Phi_{sen,rxn}$.

From the experimental results in Fig. S8 for samples containing only the sensitizer **1**, the initial and final absorbance values at 405 nm are known for each solvent. Eqn (S5) describes the temporal decrease of $A$ under the continuous incidence of $G_{ph}$ photons to the sample. This differential equation was analytically solved using the online mathematical service of Wolfram|Alpha.[S13] Finally, by applying the known parameters, the values of $\Phi_{sen,rxn}$ were calculated to be $2.7\times10^{-3}$ (hexane), $2.3\times10^{-3}$ (ethyl acetate), $8.7\times10^{-3}$ (toluene), $7.3\times10^{-4}$ (acetonitrile), $1.1\times10^{-2}$ (DMF), and $4.6\times10^{-3}$ (D-limonene), as plotted in Fig. 3e of the main text.

## 14. Photoirradiation-induced changes of optical absorption spectra of samples containing both the sensitizer and emitter in different solvents

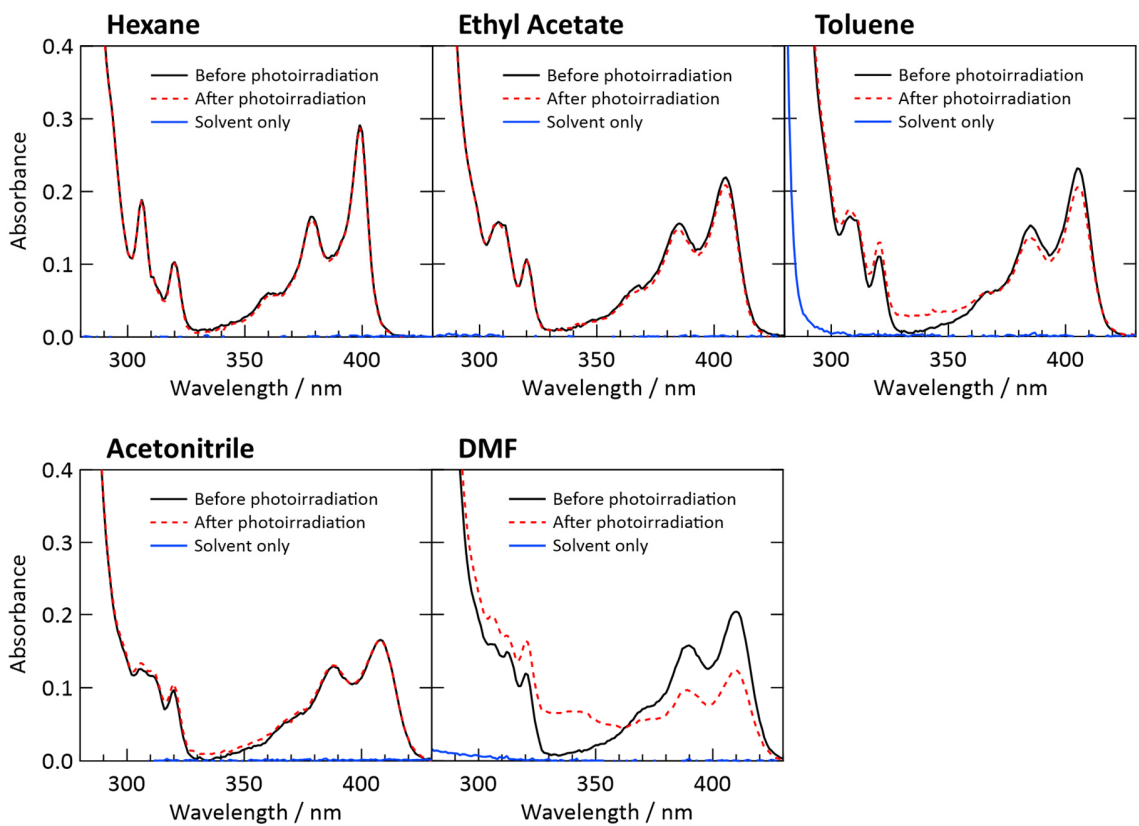

**Figure S10.** Comparison of optical absorption spectra of samples containing both sensitizer **1** and emitter **2** before and after irradiation with 405-nm light. These experiments were carried out using the experimental setup and conditions described in Section 5 of this Supplementary Information. The results for hexane are the same as those shown in Fig. 4a of the main text.

## 15. Temporal changes of fluorescence spectra of the sensitizer in the presence of the emitter during photoirradiation in different solvents

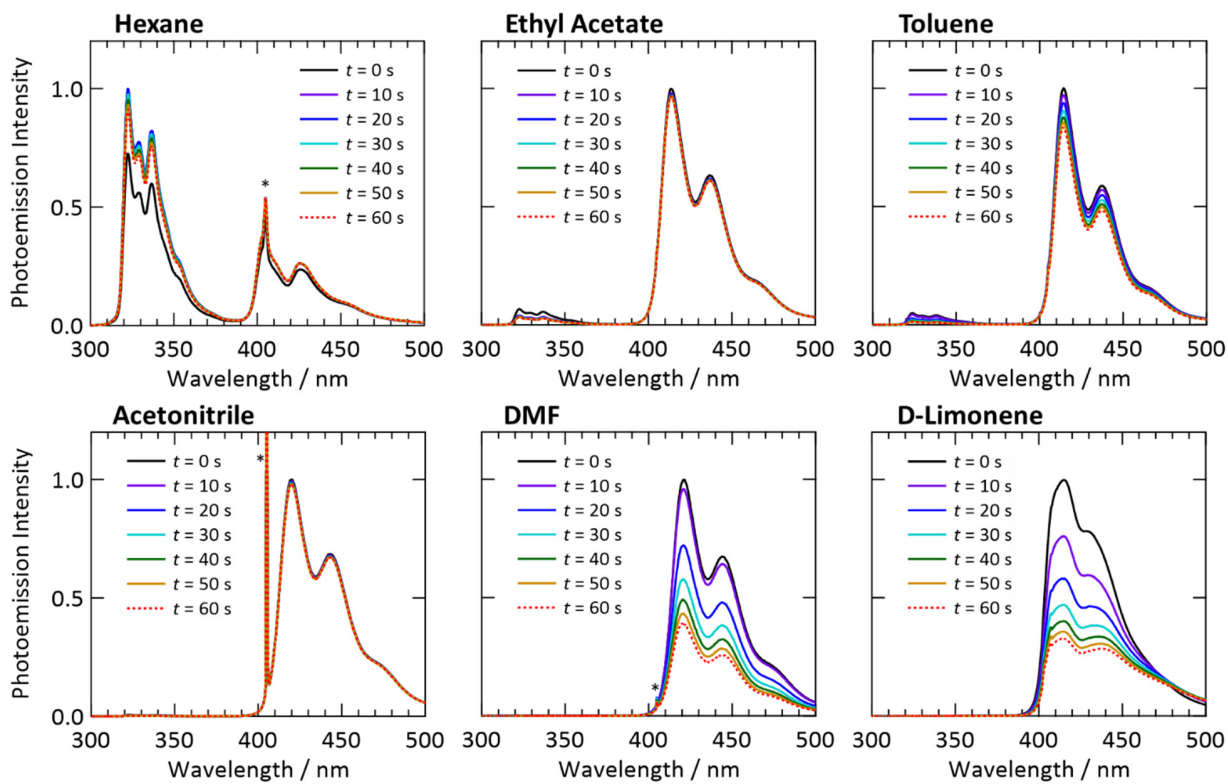

**Figure S11.** Temporal decay of the fluorescence spectrum of sensitizer **1** acquired under continuous irradiation of 405-nm laser light for samples also containing emitter **2** sealed in glass capillaries. Asterisks indicate peaks from the laser light at 405 nm. See also Fig. 4b in the main text for the spectra of the sample with hexane in a different time range of 0–80 min.

## 16. Effect of photoirradiation on the triplet lifetime of the emitter

Here, to confirm the postulation of our theoretical model described in the main text, the photoirradiation-induced generation of quenching species is investigated. To do this, we used the experimental setup and photoirradiation conditions described in Section 5 of this Supplementary Information to controllably induce photodegradation of samples before measuring triplet lifetimes.

We measured and compared the triplet lifetimes ($\tau_T$) of the emitter **2** in three samples prepared by different methods described below. All these samples used hexane, which is the representative solvent in this report. $\tau_T$ was obtained by doubling the single-exponential decay time constant of the UC emission ($\tau_{UC}$) acquired with a weak pulsed excitation where TTA is not a dominant process of triplet depopulation; i.e., $\tau_T \cong 2\tau_{UC}$.[S2] The measurements were carried out using nanosecond light pulses as described in the Experimental section of the main text.

The first sample was a normal (fresh) sample without prior photoirradiation, deaerated by FPT cycles and sealed in a glass capillary. The UC emission decay curve of this samples is indicated by black dots in Fig. S12 and its $\tau_T$ was found to be 114 µs. The second sample (control sample #1) was prepared by the following procedure. A solution containing only the sensitizer was deaerated by FPT cycles and then photoirradiated using the setup in Fig. S4. The fresh emitter was dissolved in the solution and then it was deaerated again by FPT cycles before being sealed in a glass capillary. The decay curve for control sample #1 is shown by blue dots in Fig. S12, exhibiting $\tau_T$ of 12.5 µs. The third sample (control sample #2) was prepared by photoirradiation of the normal deaerated sample containing both the sensitizer and emitter first, and then deaerated again by FPT cycles before being sealed into a glass capillary. The emission decay curve for control sample #2 is indicated by green dots in Fig. S12, showing $\tau_T$ of 63.8 µs. These results reveal that photoirradiation shortened $\tau_T$ of the emitter, which supports our postulation in the proposed model that photoirradiation generates species that quench the triplet species in the sample.

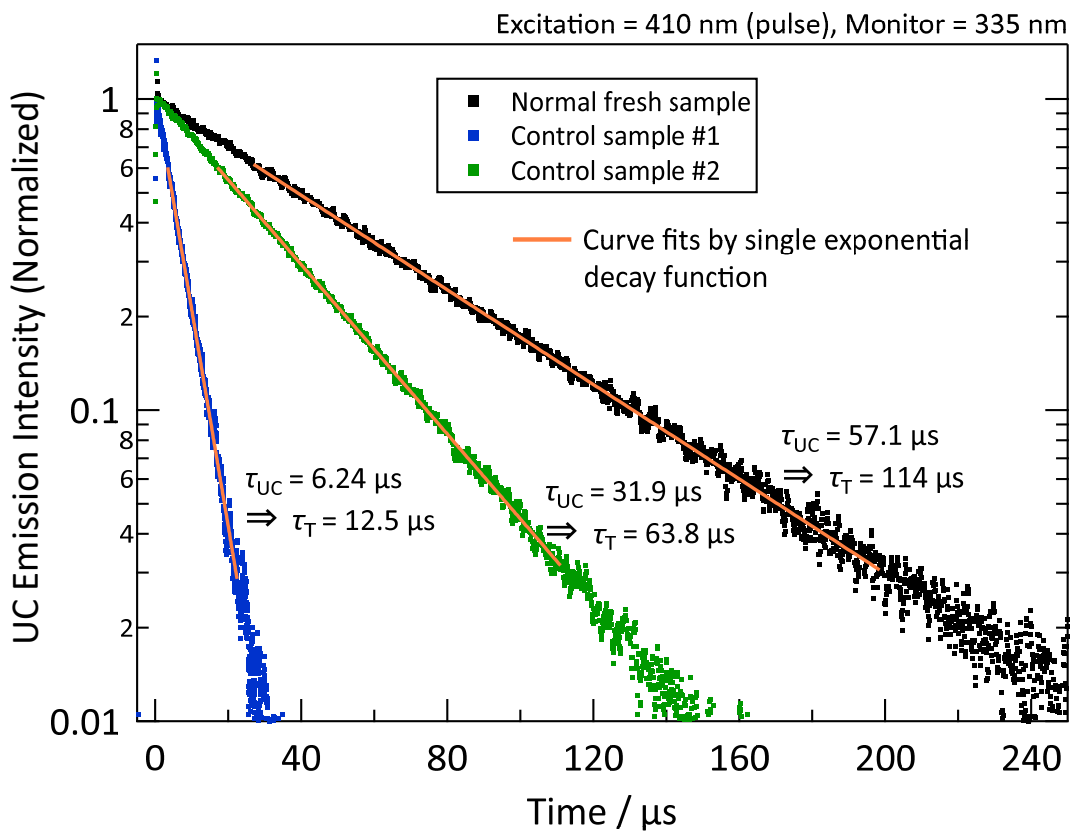

**Figure S12.** UC emission decay curves acquired for three samples prepared by different methods, which are the normal deaerated sample with fresh sensitizer and emitter (black dots), the sample prepared using the photoirradiated sensitizer solution to which fresh emitter was added and deaerated again (blue dots), and the sample first photoirradiated in the presence of both sensitizer and emitter and then deaerated again (green dots). These intensity decay curves were acquired using weak pulsed excitation at 410 nm and monitored at 335 nm. All these curves were fitted well by single-exponential decay functions, as shown by the orange lines. Determined values of $\tau_{UC}$ and $\tau_T$ are shown near each curve.

## 17. Calculation details of our theoretical model

Here we describe in detail the method used to calculate the temporal UC emission curves, examples of which are shown in Fig. 4c of the main text, from the results of our kinetic model

$$k_{\text{diff}}[E^*]^2 + \{k_T + k_{\text{diff}}([\varepsilon] + [Q])\}[E^*] - \left(\frac{[E]}{[E]+[Q]+[\varepsilon]}\right)\Gamma = 0, \qquad (S7)$$

which is eqn (10) in the main text. In eqn (S7), [E*] is the concentration of the triplet emitter; hereafter, we use the symbol $x$ in place of [E*]. Because eqn (S7) is a quadratic equation, it can be solved as

$$x = \frac{-k_\text{T}-k_\text{diff}([\varepsilon]+[Q])+\sqrt{\{k_\text{T}+k_\text{diff}([\varepsilon]+[Q])\}^2+4k_\text{diff}\Gamma\{[E]/([E]+[Q]+[\varepsilon])\}}}{2k_\text{diff}}. \tag{S8}$$

Here, the sign just before the square-root term in the numerator must be '+' to be physically valid. The magnitude of the UC emission intensity $I_\text{UC}$ at time $t$ ($I_\text{UC}(t)$), the determination of which is the purpose of this analysis, is obtained by

$$I_\text{UC}(t) = \alpha x^2, \tag{S9}$$

where $\alpha$ is an instrumental constant that can be later eliminated by appropriate normalization. The values of $k_\text{T}$ and $k_\text{diff}$ in eqn (S8) were determined from the time-resolved UC emission measurements (e.g., Fig. S11) and eqn (1) in the main text, respectively.

Once we calculate the temporal progressions of [E], [Q], and [$\varepsilon$] after the onset of photoirradiation for $t > 0$, the function $I_\text{UC}(t)$ can be obtained from eqn (S9). The initial values (at $t = 0$) for [E], [Q], and [$\varepsilon$] are $2\times10^{-3}$ ($\equiv$ [E$_0$]), Q$_0$, and 0 M, respectively. Here, Q$_0$ is an unknown constant that will be treated as an adjustable parameter in the later computation. The time progressions of these parameters are expressed by the following equations:

$$[E] = [E_0] - k_\text{emi,rxn} \int_{s=0}^{s=t} x\,ds \tag{S10}$$

$$[Q] = [Q_0] - \Phi_\text{Q,rxn} k_\text{diff} \int_{s=0}^{s=t} [Q]x\,ds \tag{S11}$$

$$[\varepsilon] = [E_0] - [E] - \Phi_{\varepsilon,\text{rxn}} k_\text{diff} \int_{s=0}^{s=t} [\varepsilon]x\,ds \tag{S12}$$

These integral equations can readily be computed by iterating numerical loops in which an infinitesimal time step $\Delta t$ is taken in each loop to calculate the temporal evolution for $t \rightarrow t + \Delta t$. In the actual computation, we introduced an additional variable [$\varepsilon_{\text{disappear}}$], which is the cumulative amount of species $\varepsilon$ deactivated by the process described by eqn (4) in the main text. Overall, the set of numerical relations used for the computation is:

$$[\text{E}(t + \Delta t)] = [\text{E}(t)] - k_{\text{emi,rxn}} x \Delta t \tag{S13}$$

$$[\text{Q}(t + \Delta t)] = [\text{Q}(t)] - \Phi_{\text{Q,rxn}} k_{\text{diff}} [\text{Q}(t)] x \Delta t \tag{S14}$$

$$[\varepsilon_{\text{disappear}}(t + \Delta t)] = [\varepsilon_{\text{disappear}}(t)] + \Phi_{\varepsilon,\text{rxn}} k_{\text{diff}} [\varepsilon(t)] x \Delta t \tag{S15}$$

$$[\varepsilon(t + \Delta t)] = [\text{E}_0] - [\text{E}(t + \Delta t)] - [\varepsilon_{\text{disappear}}(t + \Delta t)] \tag{S16}$$

By iterating the numerical loop while increasing the time by $\Delta t$ for each loop, the values of [E($t$)], [Q($t$)], and [$\varepsilon$($t$)] are obtained, from which the temporal curve of $I_{\text{UC}}(t)$ is generated. In the computation, the generated temporal curve was fitted to the experimentally acquired curve by treating Q$_0$, $k_{\text{emi,rxn}}$, $\Phi_{\text{Q,rxn}}$, and $\Phi_{\text{Q,rxn}}$ as adjustable parameters; the values of $k_{\text{emi,rxn}}$ in Fig. 4d of the main text were obtained from this fitting procedure. As mentioned in the main text, the fittings yielded Q$_0$ of $5 \times 10^{-4}$ M or lower in this study, which is equivalent to a molar fraction of 0.005% or lower. This is a trace amount and thus does not contradict the certified purities of the solvents (cf. Table S1).

## 18. Plots of $k_{\text{emi,rxn}}$ against ionization energy and electron affinity

Similar to Section 12 of this Supplementary Information, in Fig. S13 below, we plotted $k_{\text{emi,rxn}}$ against the difference between the IEs of **2** and the solvents (left) and that between the EAs of **2** and the solvents (right). As seen, $k_{\text{emi,rxn}}$ is correlated with the difference of EAs, whereas no correlation of $k_{\text{emi,rxn}}$ with the difference of IEs is found, supporting the results in Fig. 4d of the main text.

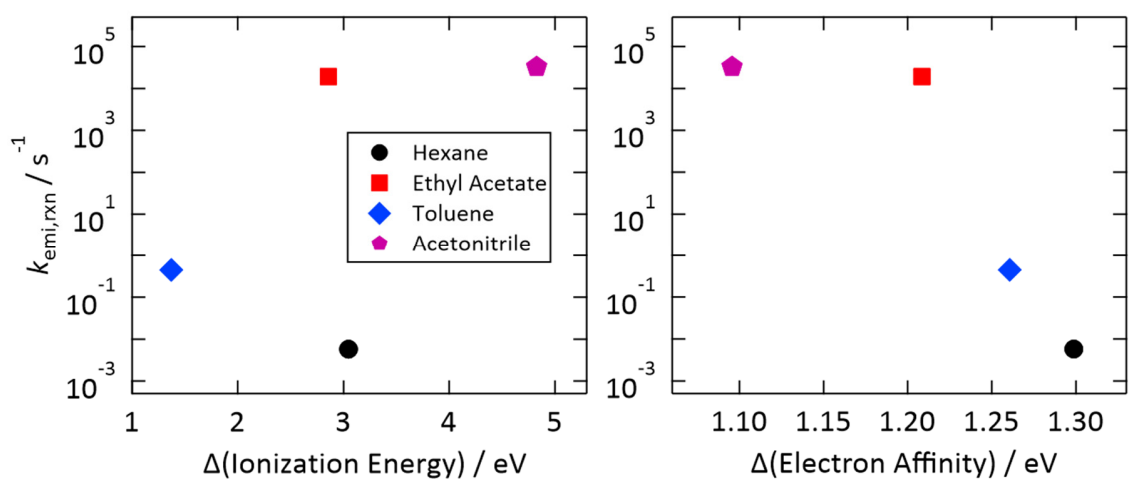

**Figure S13.** Reaction rates between the $T_1$ state of **2** and the solvents obtained from the fittings shown in Fig. 4c of the main text plotted against the difference between the ionization energies of **2** and the solvents (left) and the difference between the electron affinities of **2** and the solvents (right).